\documentclass{bmcart}
\usepackage[utf8]{inputenc} 
\usepackage{graphicx} 
\usepackage{abbreviations} 
\usepackage{natbib} 
\usepackage{amssymb}

\startlocaldefs
\newcommand{\bmath}[1]{\mathbf{#1}}

\newcommand{\bv}{\bmath{v}}

\newcommand{\bg}{\bmath{g}}

\newcommand{\bB}{\bmath{B}}
\newcommand{\bE}{\bmath{E}}

\newcommand{\text}[1]{\quad\mbox{#1}\quad}
\newcommand{\spr}[2]{\bmath{#1} \!\cdot\! \bmath{#2}}
\newcommand{\vpr}[2]{\bmath{#1} \!\times\! \bmath{#2}}
\newcommand{\vdiv}[1]{\spr{\nabla}{#1}}

\newcommand{\vcurl}[1]{\vpr{\nabla}{#1}}

\newcommand{\oder}[2]{\frac{d #1}{d #2}}

\newcommand{\Pd}[1]{\partial_{#1}}

\newcommand{\Cd}[1]{\nabla_{#1}}

\newcommand{\sub}[1]{_{\mbox{\tiny #1}}}
\newcommand{\beq}{\begin{equation}}
\newcommand{\eeq}{\end{equation}}

\newcommand{\fracp}[2]{\left(\frac{#1}{#2}\right)}
\endlocaldefs

\begin{document}
\begin{frontmatter}

\begin{fmbox}
\dochead{Research}

\title{Stationary Relativistic Jets}

\author[
   addressref={aff1,aff2},
   email={s.s.komissarov@leeds.ac.uk}
]{\inits{SSK}\fnm{Serguei S} \snm{Komissarov}}
\author[
   addressref={aff1,aff3},
   email={o.porth@leeds.ac.uk}
]{\inits{OP}\fnm{Oliver} \snm{Porth}}
\author[
   addressref={aff2},
   email={lyutikov@purdue.edu}
]{\inits{ML}\fnm{Maxim} \snm{Lyutikov}}

\address[id=aff1]{
  \orgname{School of Mathematics, University of Leeds}, 
  \postcode{LS29JT}    
  \city{Leeds},      
  \cny{UK}       
}
\address[id=aff2]{
  \orgname{Department of Physics and Astronomy, Purdue University}, 
  \postcode{47907-2036}                                
  \city{West Lafayette},                              
  \cny{USA}                                    
}
\address[id=aff3]{
  \orgname{Centre for mathematical Plasma Astrophysics, Department of Mathematics, KU Leuven}, 
  \street{Celestijnenlaan 200B},      
  \postcode{3001}                                
  \city{Leuven},                              
  \cny{Belgium}                                    
}

\end{fmbox}

\begin{abstractbox}

\begin{abstract} 
In this paper we describe a simple numerical approach which allows to 
study the structure of steady-state axisymmetric relativistic jets using 
one-dimensional time-dependent simulations. It is based on the fact that 
for narrow jets with $v_z\approx c$ the steady-state equations of 
relativistic magnetohydrodynamics can be accurately approximated by the 
one-dimensional time-dependent equations after the substitution $z=ct$. 
Since only the time-dependent codes are now publicly available this 
is a valuable and efficient alternative to the development of a high-specialised
code for the time-independent equations. The approach is also much cheaper and 
more robust compared to the relaxation method. We tested this technique against numerical 
and analytical solutions found in literature as well as solutions we obtained 
using the relaxation method and found it sufficiently accurate.    
In the process, we discovered the reason for the failure of the self-similar 
analytical model of the jet reconfinement in relatively flat atmospheres and 
elucidated the nature of radial oscillations of steady-state jets.     
\end{abstract}


\end{abstractbox}

\end{frontmatter}

\section{Introduction}
\label{sec:Intr}

Highly collimated flows of plasma from compact objects of stellar mass, like 
young stars, neutron stars and black holes, as well as supermassive black 
holes residing in the centers of active galaxies is a wide-spread phenomenon
which has been and will remain the focal point of many research programs, both 
observational and theoretical.  Some features of these cosmic jets, like moving knots, 
are best described using time-dependent fluid models. 
However, most of these jets have sufficiently regular global structure, which is 
indicative of steady production and propagation and promotes development of 
stationary models.  Such models are also easier to analyse, and they are very 
helpful in our attempts to figure out the key factors of the jet physics.  

The simplest approach to steady-state flows is to completely ignore the variation
of flow parameters across the jet. This allows to reduce the complicated system 
of non-linear partial differential equations (PDEs) describing the jet dynamics 
to a set of ordinary differential equations (ODEs) which can be integrated more 
easily \citep[e.g.][]{BR-74,ssk-94}.
A similar reduction in the dimensionality is achieved in self-similar models, 
where unknown functions depend only on a combination of independent variables 
known as a self-similar variable. This also allows to reduce the original PDEs 
to a set of ODEs \citep[e.g.][]{BP-82,VT-98}. 
While providing important test cases and useful insights, this approach is not 
sufficiently robust -- boundary and other conditions that select such exceptional 
solutions are not always present in nature. 

As it is well known to engineers working on aircraft jet engines, supersonic jets 
naturally develop quasi-periodic stationary chains of internal shocks, similar to 
what is shown in Figure~\ref{fig:fig3}. These shocks emerge as a part of the adjustment 
of the jet pressure to that of the surrounding air.  Interestingly, bright 
knots are often seen in cosmic jets and they are often interpreted as shocks 
\citep[e.g.][]{FW-85,DM-88,GM-00,ALL-10,walker-97}. Some of these knots are known 
to be travelling and they must be part of the jet's non-stationary dynamics.  
Others appear to be static and hence connected to the underlying quasi-steady-state 
structure of these cosmic jets. Quite often, the knots form  quasi-periodic chains, 
reminiscent of those seen in  aerodynamic jets. If the similarity is not accidental, 
then these knots are also related to the process of pressure adjustment. In particular, 
we expect the powerful cosmic jets to be expanding freely soon after leaving their 
central engines and to become confined by external pressure again only much later 
\citep[e.g.][]{DM-88,KF-97}. The first shock driven into the jet by the external 
pressure is called the reconfinement shock. Given the growing observational evidence of 
stationary knots in cosmic jets, there has been a increase of interest to the 
reconfinement process among theorists in recent years 
\citep[e.g][]{NS-09,Nalewajko-12,BL-07,BL-09,KB-12,KB-12a,KB-15}. 
One of the key aims of these studies was to come up with approximate analytical or 
semi-analytical solutions for the structure of steady-state jets. 

Obviously, such shocked flows cannot be described by one-dimensional (1D) and self-similar 
models, which we mentioned earlier, and more complex, at least two-dimensional (2D), 
models have to be applied instead. 
The system of steady-state equations of compressible fluid dynamics, not to mention 
magnetohydrodynamics, is already very complicated and generally 
requires numerical treatment. One of the ways of finding its solutions involves integration
of the original time-dependent equations in anticipation that if the boundary conditions are
time-independent then the time-dependent numerical solution will naturally evolve towards 
a steady-state \citep[e.g.][]{UKRCL-99,kvkb-09,TMN-08}. 
One clear advantage of this approach is that it allows to use standard codes for 
time-dependent fluid dynamics. Such codes are now well advanced and widely available. 
However, this type of the relaxation approach is characterised by slow convergence and 
hence rather expensive. 

In order to speed up the convergence, one can use other relaxation methods, which are developed
specifically for integrating steady-state equations \citep[e.g.][]{MJ-11}. They often involve   
a relaxation variable which is called ``pseudo time''.  However, this time evolution is not 
realistic but designed to drive solutions towards a steady-state in the fastest way possible.     
The only disadvantage of this approach is that it involves development of a specialised 
computer code dedicated to solving only steady-state problems. The authors are not aware 
of such codes for relativistic hydro- and magnetohydrodynamics.   

For supersonic flows, the system of steady-state equations turns out to be 
hyperbolic, with one of spatial coordinates playing the role of time \citep{Glaz-85}. 
(In the case of magnetic jets, the speed of sound is replaced with
the fast magneto-sonic speed and we classify flows as sub-, tran-, or super-sonic 
based on its value compared to the flow speed.) 
In this case, one can find steady-state solutions utilising numerical methods which were 
designed specifically for hyperbolic systems, like the method of characteristics 
or ``marching'' schemes. These methods have been used in the past in applications to 
relativistic jets \citep[e.g.][]{DM-88,WF-85,wilson-87,bowman-94,bowman-96} but
publicly available codes do not exist yet. 
Their development is as time-consuming as that of time-dependent codes whereas the 
range of applications is much more limited. This explains their current unavailability. 
Moreover, when flow becomes subsonic, even very locally, this approach fails.               

In this paper, we propose a new approach, which allows to find approximate numerical 
steady-state jet solutions rather cheaply and using widely available computer 
codes. To be more precise, we focus on highly relativistic narrow axisymmetric jets 
and show that in this regime the 2D steady-state equations of Special  
Relativistic MHD (SRMHD) are well approximated by 1D time-dependent equations 
of SRMHD. Like in the standard marching schemes, the spatial coordinate along the 
jet plays the role of time. This allows us to find steady-state structure of 
axisymmetric jets by carrying out basic 1D SRMHD simulations, which can be 
done with very high resolution even on a very basic personal computer. 
In such simulations, no special effort is needed to preserve the magnetic field 
divergence-free and the computational errors associated with multi-dimensionality 
are eliminated. As the result, more extreme conditions can be tackled. 
Here we focus only on relativistic jets, 
because of  our interest to AGN and GRB jets, but we see no reason why this approach 
cannot be applied to non-relativistic hypersonic jets as well.   
Our approach is closely related to the so-called ``frozen pulse'' approximation, 
which also utilises the similarity between the steady-state and time-dependent 
equations describing ultra-relativistic flows \cite{PN-93,VK-03a,SV-13}. In this 
approximation, the steady-state equations are used to analyse the dynamics of 
time-dependent flows. 
The similarity between 1D time-dependent models and 2D
steady-state jet solutions has been noted before, in particular in \cite{MMS-12}.

In order to study the potential of this new approach we have carried out a number of 
test simulations and compared the results obtained in this way with both 
analytical models and numerical solutions obtained with more traditional methods.     
The results are very encouraging and allow us to conclude that this method is viable and 
can be used in a wide range of astrophysical applications.   
    
\section{Approximation} 
\label{sec:approximation}

We start by writing down the time-dependent equations of Special Relativistic 
Magnetohydrodynamics (SRMHD).  In this section we use units where the speed of light $c=1$ and
the factor $1/4\pi$ does not appear in the expression for the electromagnetic energy 
density. The components of vectors and tensors are given in normalised bases.  
The evolution equations of SRMHD include the continuity equation

\begin{equation}
   \Pd{t}{\rho\Gamma} + \bmath{\nabla}\cdot (\rho\Gamma \bv)=0\,,
\label{ce}
\end{equation}
the Faraday equation 

\begin{equation}
   \Pd{t}\bB + \vcurl{E} = 0\, .
\label{Faraday}
\end{equation}
and the energy-momentum equation

\begin{equation}
   \Pd{t}{T^{t\mu}}+\nabla_j{T^{j\mu}}=0\, ,
\label{EM}
\end{equation}
where

\begin{equation}
   T^{\nu\mu}=T^{\nu\mu}\sub{hd} + T^{\nu\mu}\sub{em}
\end{equation}
is the total stress-energy-momentum tensor,

\begin{equation}
   T^{\nu\mu}\sub{hd} = wu^\nu u^\mu +pg^{\nu\mu}
\end{equation}
is stress-energy-momentum tensor of matter and the components of 
the electromagnetic stress-energy-momentum tensor are

\begin{equation}
   T^{tt}\sub{em} =(E^2+B^2)/2 \, ,
\end{equation}
\begin{equation}
   T^{ti}\sub{em} =(\vpr{E}{B})^i \, ,
\end{equation}
\begin{equation}
   T^{ij}\sub{em} =-(E^iE^j+B^iB^j)+\frac{1}{2}(E^2+B^2)g^{ij} \,.
\end{equation}
In these equations, $\bB$ and $\bE$ are the vectors of magnetic and electric fields respectively, 
$p$, $\rho$ and $w$ are the thermodynamic pressure, rest-mass density of matter and relativistic 
enthalpy of matter respectively, $\bv$ is the velocity vector, $\Gamma$ is the Lorentz factor and
$\bg$ is the metric tensor of space.
These equations are to be supplemented with Equation of State 
$w=w(\rho,p)$ and the Ohm's law of ideal MHD

\begin{equation}
\bE = -\vpr{v}{B}\, .
\label{ohm}
\end{equation}
Finally, the magnetic field is divergence-free
  
\begin{equation}
   \vdiv{B}=0\, .
\label{GaussB}
\end{equation}

In this analysis, we focus on axisymmetric jets and adopt a cylindrical coordinate system 
with the z axis coincident with the jet symmetry axis. We consider only narrow jets, so
that 

\beq
    \frac{r}{z} \ll 1 \,.
\label{cond1a}
\eeq

We also constrain ourselves with a relatively simple magnetic configurations where the 
divergence-free condition leads to  
\beq
   \frac{B^r}{B^z} \simeq \frac{r}{z} \ll 1 \,. 
\label{cond1b}
\eeq
In axisymmetry, the steady-state Faraday equation implies $E^\phi=0$. When combined with
Eq.\ref{ohm}, this result yields  

\beq
      \frac{v^r}{v^z} =  \frac{B^r}{B^z}  \ll 1 \,.
\label{cond1}
\eeq
Thus, the radial components of both the magnetic field and the velocity vectors are 
small compared to their axial components. 

We also assume that $v^\phi \ll 1$. In fact, in the case of magnetically accelerated jets,     

$$
    v^\phi \simeq (r\sub{lc}/r) 
$$ 
when $r\gg r\sub{lc}$, the radius of light cylinder (See eq.66 in \cite{kvkb-09}). 
Thus, this is a good 
approximation for astrophysical jets.  For a highly relativistic flow, 
the condition $v^z \gg v^r,v^\phi$ means 
\beq
    v^z \simeq 1 \,. 
\label{cond2}
\eeq 
Following the standard flux freezing argument, along the jet
$B^\phi/B^z \simeq (r\sub{j}/r\sub{lc})^{-1}$, where and $r\sub{j}$ is the jet radius.
(This argument does not apply to turbulent jets, which are 
non-axisymmetric and  allow non-trivial conversion of components.)
Hence one may argue that far away from the central engine 
\beq
      B^\phi \gg B^z \, .
\label{cond3}
\eeq

In order to introduce the key idea of our approach we consider first the steady-state 
continuity equation: 

\begin{equation}
   \Pd{z} (\rho\Gamma v^z) + \Cd{r} (\rho\Gamma v^r)=0\, .
\label{ce1}
\end{equation}
Using the condition (\ref{cond2}) we may replace $v^z$ with unity. This makes Eq.\ref{ce1} 
identical to the 1D time-dependent version of the continuity equation. In order to stress
this point we replace $z$  with $t$ and write:    
     
\begin{equation}
   \Pd{t} (\rho\Gamma) + \Cd{r} (\rho\Gamma v^r)=0\, .
\label{ce2}
\end{equation}
Similarly, all 2D steady-state equations can be approximated by the corresponding 
1D time-dependent equations. 

Let us show this for the equations of magnetic field. 
The 1D version of the divergence free condition reads
$$
      \Pd{r} (rB^r) = 0 \text{or} r B^r =\mbox{const}.   
$$
Thus if $B^r$ vanishes outside of the jet, which is expected when it is in direct
contact with ISM, then one has to put $B^r=0$ everywhere in the 1D model. 
As we shell see, the terms involving  
$B^r$ are sub-dominant in all other equations and hence this is a reasonable 
simplification.  Moreover, once the 1D solution is found, one can substitute the 
determined $B^z(r,z)$  into the 2D divergence free condition and solve 
it for $B^r(r,z)$. The result can then be used to verify that $B^r(r,z) \ll B^z(r,z)$.

The $\phi$ component of the Faraday equation can be written as 

\begin{equation}
    \Pd{t}B^\phi
    -rB^i\Pd{i}\fracp{v^\phi}{r} + 
     \Pd{i} (v^i B^\phi)  =0\, ,
\label{Faraday-phi}
\end{equation}
where $i={r,z}$.  In steady-state, the first term vanishes, the next two terms are of the 
order $B^z v^\phi/z$ and small compared to the last two terms, which are of the 
order $B^\phi v^z/z$. Removing these small terms we obtain the approximate steady-state 
equation  
\begin{equation}
     \Pd{z} (v^z B^\phi) + \Pd{r} (v^r B^\phi)  =0\, .
\end{equation}
Finally, we replace $v^z$ with unity, $z$ with $t$, and obtain
  
\begin{equation}
   \Pd{t} B^\phi+ \Pd{r} (v^r B^\phi) =0\, . 
\label{Faraday-phi-1d}
\end{equation}
This is indeed the 1D version of the $\phi$ component of  eq.\ref{Faraday-phi}. 
Now consider the $z$ component of the Faraday equation, 

\begin{equation}
    \Pd{t} B^z -B^i\Pd{i}v^z + 
   \frac{1}{r} \Pd{i} (r v^i B^z) = 0\, .
\label{Faraday-z}
\end{equation}
The last two terms of this equation are of the order $v^z B^z/z \simeq B^z/z$. 
On the other hand, the second and the third terms are much smaller because of the special 
status of $v^z$, which is approximately constant, and hence $B^z\Pd{z}v^z \ll B^z (v^z/z)$. 
Removing these small terms, we obtain the approximate steady-state equation 

\begin{equation}
   \Pd{z} (v^z B^z) + \frac{1}{r} \Pd{r} (r v^r B^z) =0\, . 
\end{equation}
Now once again we replace $v^z$ with unity and $z$ with $t$ to obtain 

\begin{equation}
   \Pd{t} B^z+
   \frac{1}{r} \Pd{r} (r v^r B^z) =0\, , 
\label{Faraday-z-1d}
\end{equation}
which is the 1D version of the $z$ component of  eq.\ref{Faraday-phi}.

Finally, we analyse the energy-momentum equations. These can be written as  

\begin{equation}
   \Pd{t}{T^{t\mu}} + \Pd{z}T^{z\mu} + \nabla_r{T^{r\mu}}=0\, . 
\label{EM-1}
\end{equation}
so the steady-state versions are

\begin{equation}
   \Pd{z}T^{z\mu} + \nabla_r{T^{r\mu}}=0\, . 
\end{equation}
These already have the same form as the 1D time-dependent equations, so we only need to show 
that 
\beq
T^{z\mu}\simeq T^{t\mu} \,. 
\eeq
Let us start with the hydrodynamic contribution.
First, we notice that
$$
 T^{tt}\sub{hd} = w\Gamma^2-p \simeq w\Gamma^2 \text{as} \Gamma\gg 1\,;
$$ 
$$
  T^{tz}\sub{hd} = w\Gamma^2 v^z \simeq  w\Gamma^2 \text{as} v^z\simeq 1\,.
$$
Thus, $T^{zt}\sub{hd}\simeq T^{tt}\sub{hd}$. Then we notice that 
$$
 T^{ti}\sub{hd} = w\Gamma^2v^i\,;
$$
$$
  T^{zr}\sub{hd} = w\Gamma^2 v^z v^r \simeq  w\Gamma^2 v^r\,;
$$
$$
  T^{zz}\sub{hd} = w\Gamma^2 v^z v^z \simeq  w\Gamma^2 v^z\,;
$$
Thus, $T^{zi}\sub{hd}\simeq T^{ti}\sub{hd}$. 

Now we inspect the electromagnetic contributions. First, we find good 
estimates for the components of electric field. From eq.\ref{ohm} it follows 
that   
\beq
E^r \simeq B^\phi 
\label{Er}
\eeq
and 
$$
E^z=B^rv^\phi-B^\phi v^r \ll E^r\, . 
$$  
In fact, it is easy to show that 
\beq
  E^z \simeq -B^\phi v^r \,. 
\label{Ez}
\eeq
Indeed, for magnetically accelerated jets $B^\phi \simeq \Omega r B^z$ (e.g. \cite{kvkb-09})
for $r\gg r\sub{lc}$. Hence
$$
     v^r B^\phi \simeq v^r\Omega r B^z \simeq (r/r\sub{lc}) B^r \gg B^r \gg B^rv^\phi \,.
$$
Using these estimates we find that
$$
   T^{tt}\sub{em} = \frac{1}{2}(E^2 + B^2) \simeq B_\phi^2 \,;
$$
$$
   T^{tz}\sub{em} = (\vpr{E}{B})^z \simeq E^r B^\phi \simeq B_\phi^2 \,,
$$
and hence $T^{zt}\sub{em}\simeq T^{tt}\sub{em}$. Moreover,  
 
$$
   T^{zz}\sub{em} = -(E_z^2+B_z^2) +\frac{E^2+B^2}{2} \simeq 
   B_\phi^2 \, ,
$$
and hence $T^{zz}\sub{em}\simeq T^{tz}\sub{em}$ as well. 
Next we show that $T^{z\phi}\sub{em}\simeq T^{t\phi}\sub{em}$. 
Indeed, 
$$
   T^{t\phi}\sub{em} = E^zB^r - E^rB^z \simeq - E^rB^z   \, ,
$$
and 
$$
   T^{z\phi}\sub{em} = -(E^z E^\phi+B^z B^\phi)  
\simeq  -B^z E^r \,.  
$$
Finally, we show that $T^{zr}\sub{em}\simeq T^{tr}\sub{em}$. First,  we find 
straight away that 
$$
     T^{tr}\sub{em} = -E^z B^\phi  \text{and}  
     T^{zr}\sub{em} = -(E^zE^r+B^zB^r)   \,.
$$ 
Since $E^zE^r \simeq E^z B^\phi$, we only need to show that $B^zB^r$ is 
significantly smaller compared to these terms. This is indeed the case as 
$ B^zB^r \simeq v^r B_z^2$ whereas using Eqs.\ref{Er} and \ref{Ez} we obtain 
$ E^zE^r \simeq v_r B_\phi^2 \gg v^r B_z^2$. 

Thus, within our approximation the steady-state 2D equation of energy-momentum
reduces to 

\begin{equation}
   \Pd{t}{T^{t\mu}} +  \nabla_r{T^{r\mu}}=0\, , 
\label{EM-2}
\end{equation}
which is the 1D time-dependent energy-momentum equation. 
  
Given that in relativistic fluid dynamics small differences between  
the magnitudes of energy and momentum may result in huge variations of Lorentz
factor and even lead to inconsistency, one could feel uneasy about the 
approximations we make.  However, the final result is {\it exactly} the system of 
1D time-dependent SRMHD and this means that self-consistency is not compromised. 
For example, the flow speed will not exceed the speed of light because of the 
errors of our approximation.

Our approach is similar to ``marching'' -- we compute solution for a downstream jet 
cross-section using only the previously found solutions for upstream cross-sections. 
Strictly speaking, this requires the flow to 
be super-sonic for unmagnetised jets and super-fast-magnetosonic for magnetised ones 
\citep{WF-85,DP-93}.  However, in our derivations we never had to utilise this condition. 
This suggests that it is not required when we wish to find only approximate solutions. 
For example, one may argue that the fact that information can propagate upstream does not
necessarily imply that this always has a strong effect on the flow -- the upstream-propagating 
waves could be rather weak. If so, we may still apply our method to jets where the supersonic 
condition is not fully satisfied, but we always need to check that the conditions 
(\ref{cond1a}-\ref{cond3}) of our approximation hold for obtained solutions.

\section{Numerical Implementation}
\label{sec:num-implement}

The analysis of Section~\ref{sec:approximation} shows that as long as they are applied to 
narrow jets with high Lorentz factor, the axisymmetric steady-state equations of SRMHD are 
very close to 1D time-dependent equations of SRMHD in cylindrical geometry. This suggests that 
it may be possible to use time-dependent simulations with 1D SRMHD codes to study the 2D structure
of steady-state jet solutions. However in order to be able to do this, we also need to find 
a way of accommodating the 2D boundary conditions of steady-state problems in such 
simulations.

For 2D supersonic flows we need to fix all flow parameters at the jet inlet and impose  
some conditions at the jet boundary, consistent with it being a stationary contact wave. 
No boundary conditions are needed for the outlet boundary - its flow parameters are 
part of the solution. In the corresponding 1D problem, the 2D boundary conditions 
at the inlet boundary simply become the initial conditions of the 1D Cauchy problem. 
The final 1D solution corresponds to the slice of the 2D solution at the outlet boundary. 
As to the contact discontinuity at the 2D jet boundary, the situation is not that 
trivial. 

Suppose that the total pressure at this boundary is a function of $z$, 
$p=p\sub{b}(z)$. When we replace $z$ with $t$ this becomes $p=p_b(t)$. 
Thus we need somehow to impose time-dependent boundary conditions. 
In the  simulations presented below, the following approach was utilised:  
1) we extend the computational domain so that it includes the external gas, 
2) we track the point separating the jet from the external gas and 
3) we reset the external gas parameters according to the prescribed functions 
of time every computational time step. 

In order to locate the boundary separating the jet from the external gas, we 
employ a simplified version of the level-set method \citep{OS-88,SS-03}. 
Namely, we introduce the passive scalar $\tau$, which satisfies the advection equation 
\beq
\Pd{t}(\Gamma \rho \tau) + \frac{1}{r} \Pd{r}(r\Gamma v^r \rho \tau ) = 0 \, .
\eeq
The initial solution has a smooth distribution of this scalar 
\beq
 \tau=\frac{1}{2}\left(1 - \tanh\frac{r-r_j}{\Delta} \right)\,,
\eeq
with the value $\tau=0.5$ corresponding to the jet boundary (In the test simulations, we
used $\Delta=0.3r_j$.). During the simulations, the condition $\tau<0.5$ was used to 
identify the external gas. 

After the reset, the 1D jet boundary is no longer a contact but a more general 
discontinuity. In particular, the jet plasma will generally have radial velocity 
component. If it is positive, but in the external gas it is set to zero, 
then a shock wave will launched into the jet when this discontinuity is resolved.    
If it is negative, then this will be a rarefaction wave. On the one hand, this reflects
how the information about changing environment is communicated to the interior of a 
steady-state jet. On the other hand, in 1D simulations the strength of the emitted 
wave depends on the external density -- higher density, and hence lower temperature, 
will result in stronger waves moving into the jet. This is obviously not so  
for 2D steady-state jets, which react only to the external pressure. 
Thus additional measures need to be undertaken. First, in order to negate the effect of 
the radial velocity jump at the jet boundary, the radial velocity of the 
external gas is reset not to zero but to its value at the last jet cell. Second, in order
reduce the role of the external gas inertia, it helps to set its density to a low 
value, so that its sound speed becomes relativistic. Although we have not 
tried this, one could set the polytropic index of the external gas to $\Gamma=2$, 
which would make the sound speed of ultra-relativistically hot gas equal to the 
speed of light.

\section{Examples}

\subsection{Bowman's jet}
\label{sec:bowman}

To test the validity of our approach, we first use our method 
to reproduce the numerical steady-state solutions for supersonic unmagnetised jets obtained 
by Bowman \cite[][B94]{bowman-94} using the marching scheme described in \cite{wilson-87}. 
In this study pressure-matched uniform jets with zero opening angle are injected 
into an atmosphere with the pressure distribution

\beq
p(z) = p_0 \left[ \left( \frac{z}{z_{\rm s}}\right)^{-2} + 
     \left(1-\frac{z_{\rm s}}{z}\right) 
     \left( \frac{z_{\rm s}}{z_{\rm c}}  \right)^2\right] 
\label{eq:bowman_p1}
\eeq
with $z_{\rm s}=10$, $z_{\rm c}=50$. According to this equation, the external pressure
initially decreases almost as fast as $\propto z^{-2}$ but at $z>z\sub{c}$ becomes uniform. 
The initial jet radius $r_0=1$ and the injection 
nozzle is located at $z=z\sub{s}$. The equation of state is that of Synge \cite{S-57} 
for an electron-proton plasma. 
The initial jet pressure $p\sub{j}=p_0$.  For the comparison 
we selected the model with the Mach number $M\sub{j}=15$ and the initial temperature 
$T\sub{j}=\sqrt{10}\times 10^{13}\rm K$. At such a high temperature the EOS of 
electro-proton plasma is almost the same as that of the pure proton gas. The latter was 
used in our simulations.

\begin{figure}
\includegraphics[width=9cm]{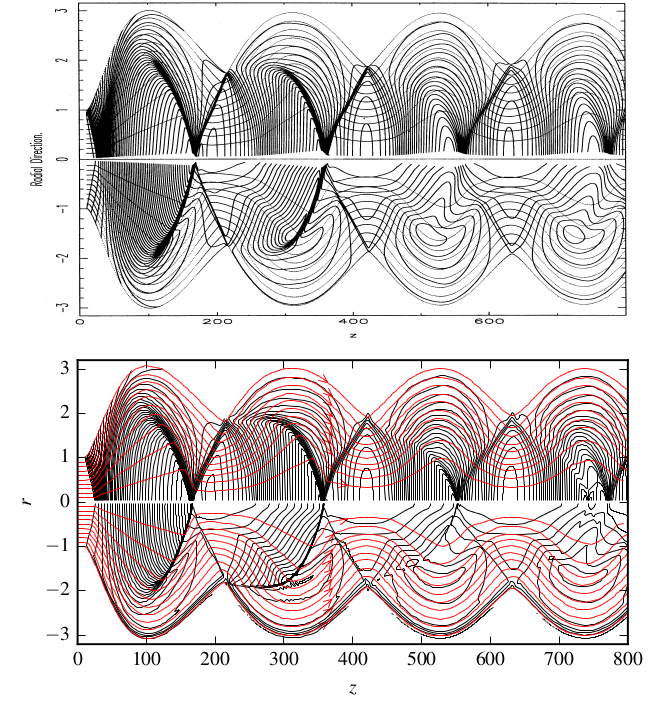}\\
\caption{Reconfinement of the $M_{\rm j}=15,\ T_{\rm j}=\sqrt{10}\times 10^{13}\rm K$ jet.
The top panel is a reproduction of Figure 3 from B94. The bottom panel shows the solution 
obtained with our method. In each panel, the top halves show 50 pressure contours 
(spaced by the factor of $1.18$) and the bottom halves show the temperature 
parameter $\tau\equiv \rho h/(\rho h - p)$ in 50 contours 
(spaced by the factor of $1.003$). The light grey lines are streamlines.  }
\label{fig:fig3}
\end{figure}

Bowman's solution is shown in the top part of figure~\ref{fig:fig3}. As the external pressure 
decreases rapidly, the jet quickly becomes under-expanded and enters the phase of 
almost free expansion. When it enters the outer region of constant pressure it becomes 
over-expanded and a reconfinement shock is pushed towards its axis, where it gets reflected.
Gas passed though these two shocks becomes hot and its pressure rises. As a result, the jet 
becomes somewhat under-expanded again and begins to expand for the second time. Then it becomes 
over-expanded again and another shock is pushed into the jet and so on. 

In the bottom part of this figure, we show the results of our 1D
simulations for this jet using exactly the same visualisation
technique as in the original paper. The agreement between the two solutions is quite
remarkable.  A very good match for the maximal radial extension and
the oscillation-length of the jet is obtained.  The successive
reconfinement shocks are somewhat sharper than in B94, most likely due
to the application of a shock-capturing scheme.
We checked our approach against other numerical models of B94 as well.   
In all models, the results for profile of jet radius and Mach number are in
good agreement.  Noticeable but still minor differences arise only for the 
colder models, most likely due to the different equation of state used in our 
simulations.

\subsection{Self-similar models of jet reconfinement}
\label{sec:validation_hot}

The problem of reconfinement of initially free-expanding steady-state jets is quite 
important and a number of authors have tried to find simple analytic of semi-analytic
solutions. Falle \cite{F-91} and Komissarov \& Falle \cite{KF-97} used the Kompaneets approximation, which 
assumes that the gas pressure immediately downstream of the reconfinement shock is equal to 
the external pressure at the same distance, to derive a simple ODE for the shock 
radius. Assuming particular flow profiles in the shocked layer, one can also determine 
the location of the jet boundary \citep[e.g.][]{BL-07}. The Kompaneets approximation is 
accurate only for very narrow jets. To improve on it, one also has to take into account the  
variation of the gas pressure across the shocked layer \citep{NS-09}. In our second test, 
we compare our results with the semi-analytical model by  \citep[][thereafter KBB12]{KB-12}, 
who assumed self-similarity of the flow in this layer. This assumption is 
more suitable for the case where the reconfinement shock never reaches the jet axis, 
because otherwise the distance where this occurs sets a characteristic length scale.  

%
\begin{figure*}
\begin{center}
\includegraphics[height=8cm]{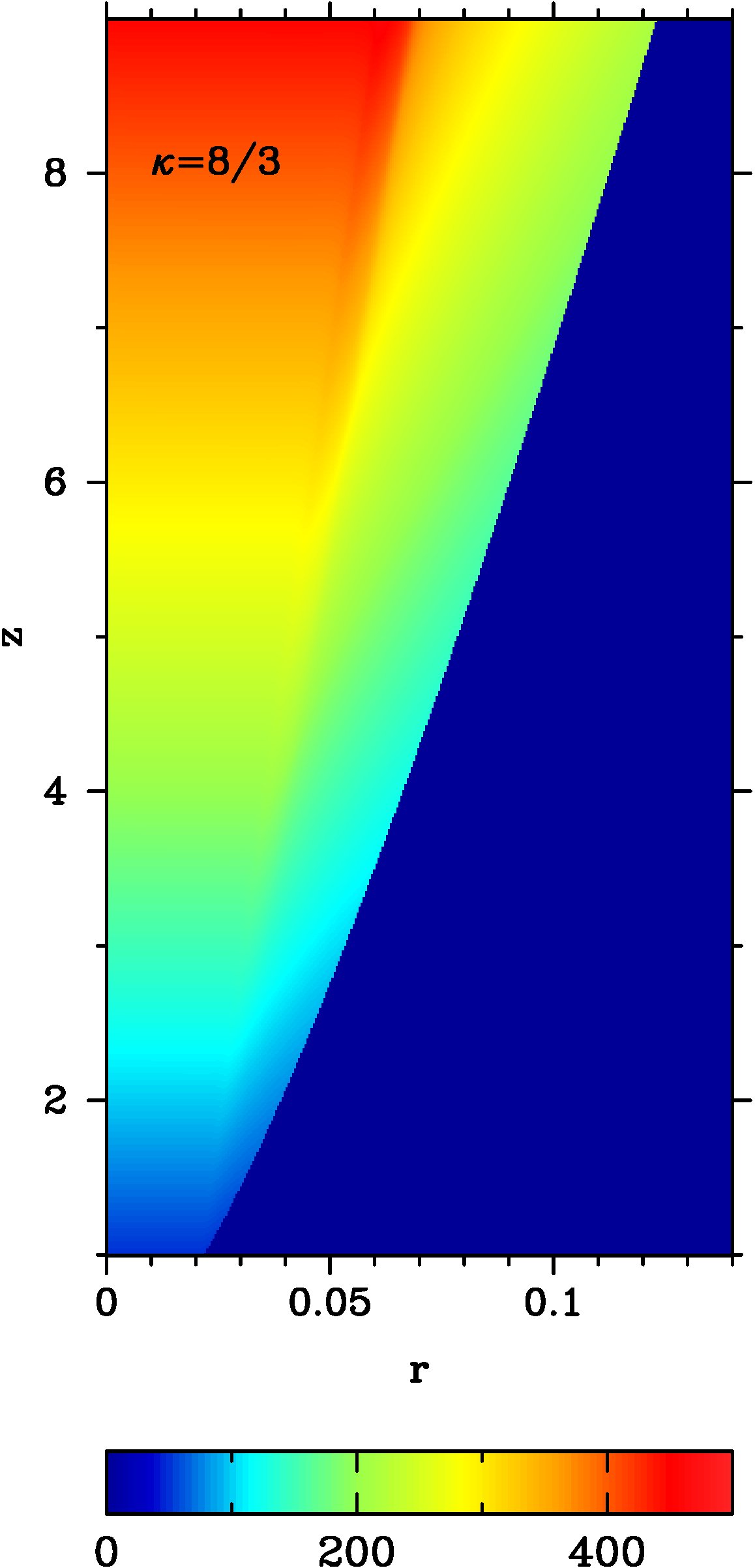}
\includegraphics[height=8cm]{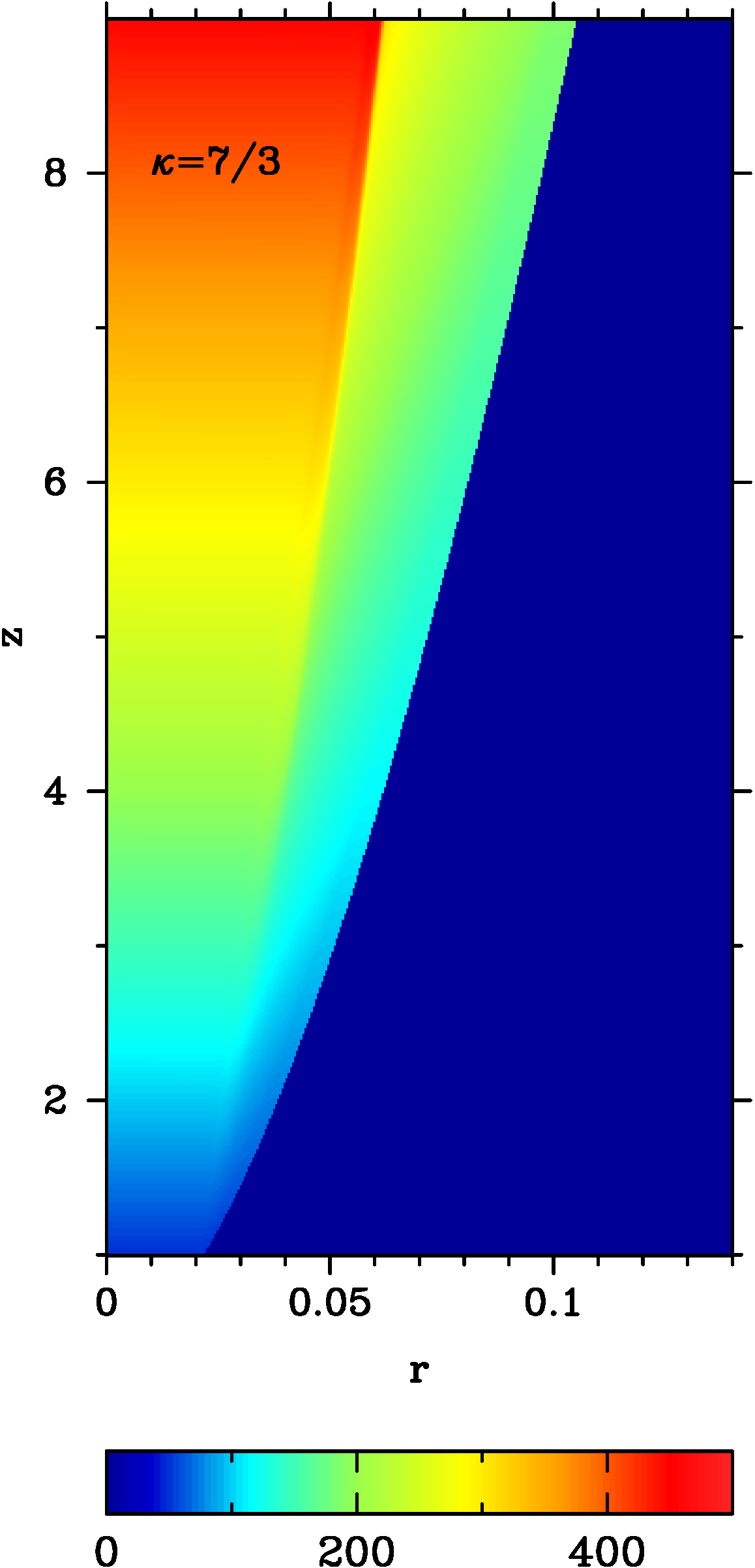}
\includegraphics[height=8cm]{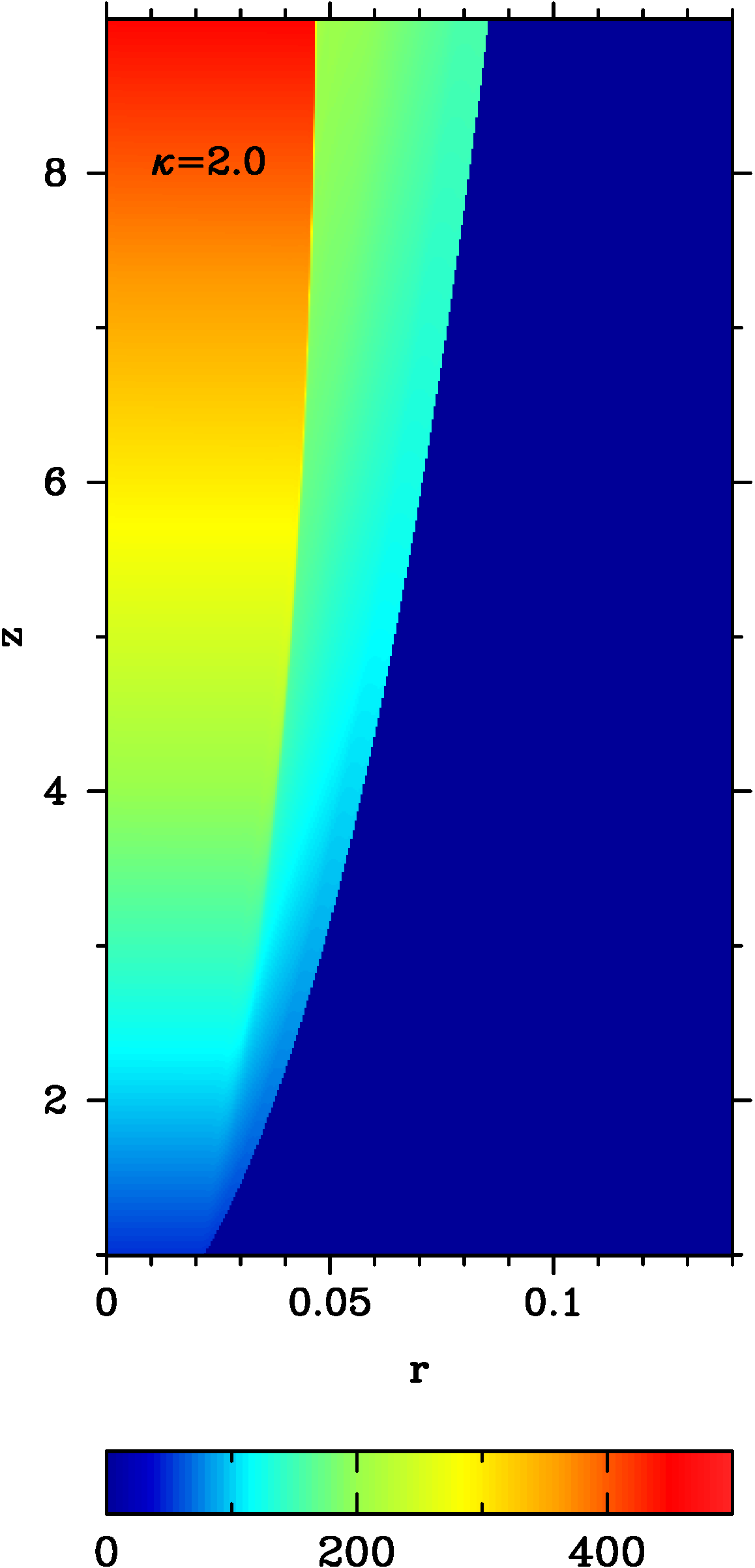}
\caption{
Ultra-relativistically hot jets \citep{KB-12,KB-15} in power-law 
atmospheres with $\kappa=8/3,7/3$ and 2 (from left to right). The colour-coded 
images show the distribution of the Lorentz factor. The initial Lorentz factor is 
$\Gamma_0=50$ and opening angle $\theta_0=1/\Gamma_0$.}
\label{fig:hot1}
\end{center}
\end{figure*}
%
%
\begin{figure*}
\begin{center}
\includegraphics[height=8cm]{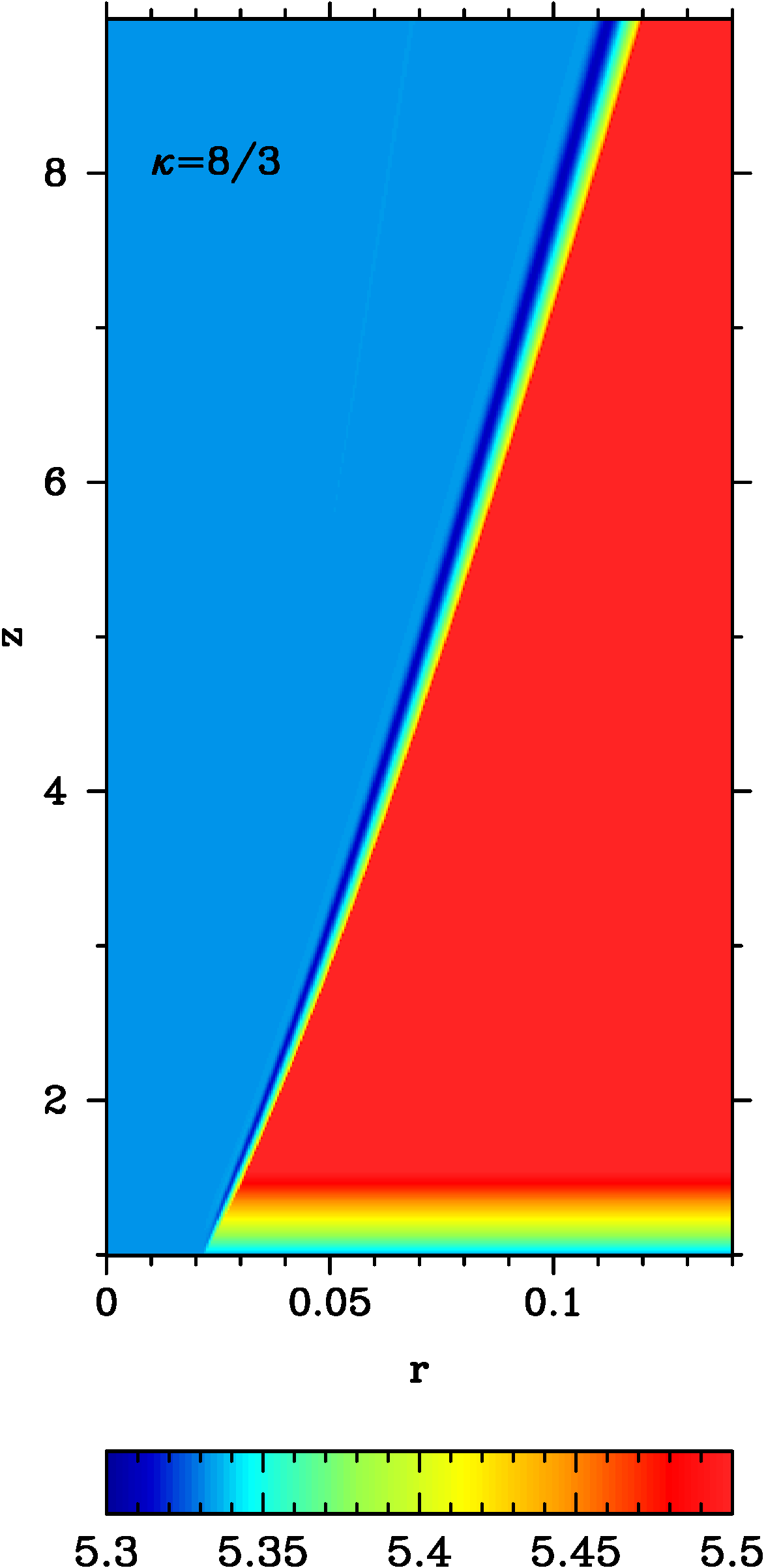}
\includegraphics[height=8cm]{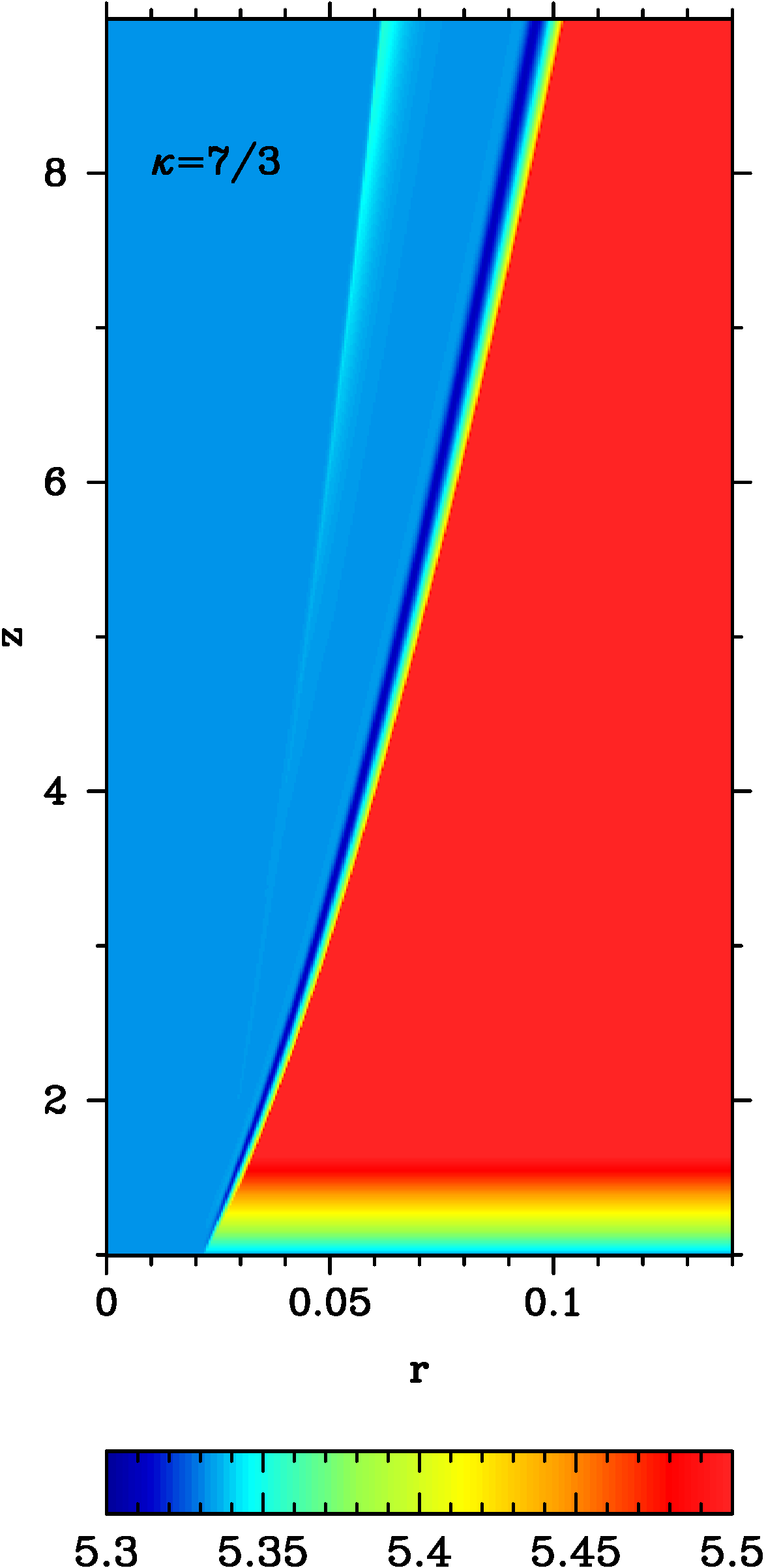}
\includegraphics[height=8cm]{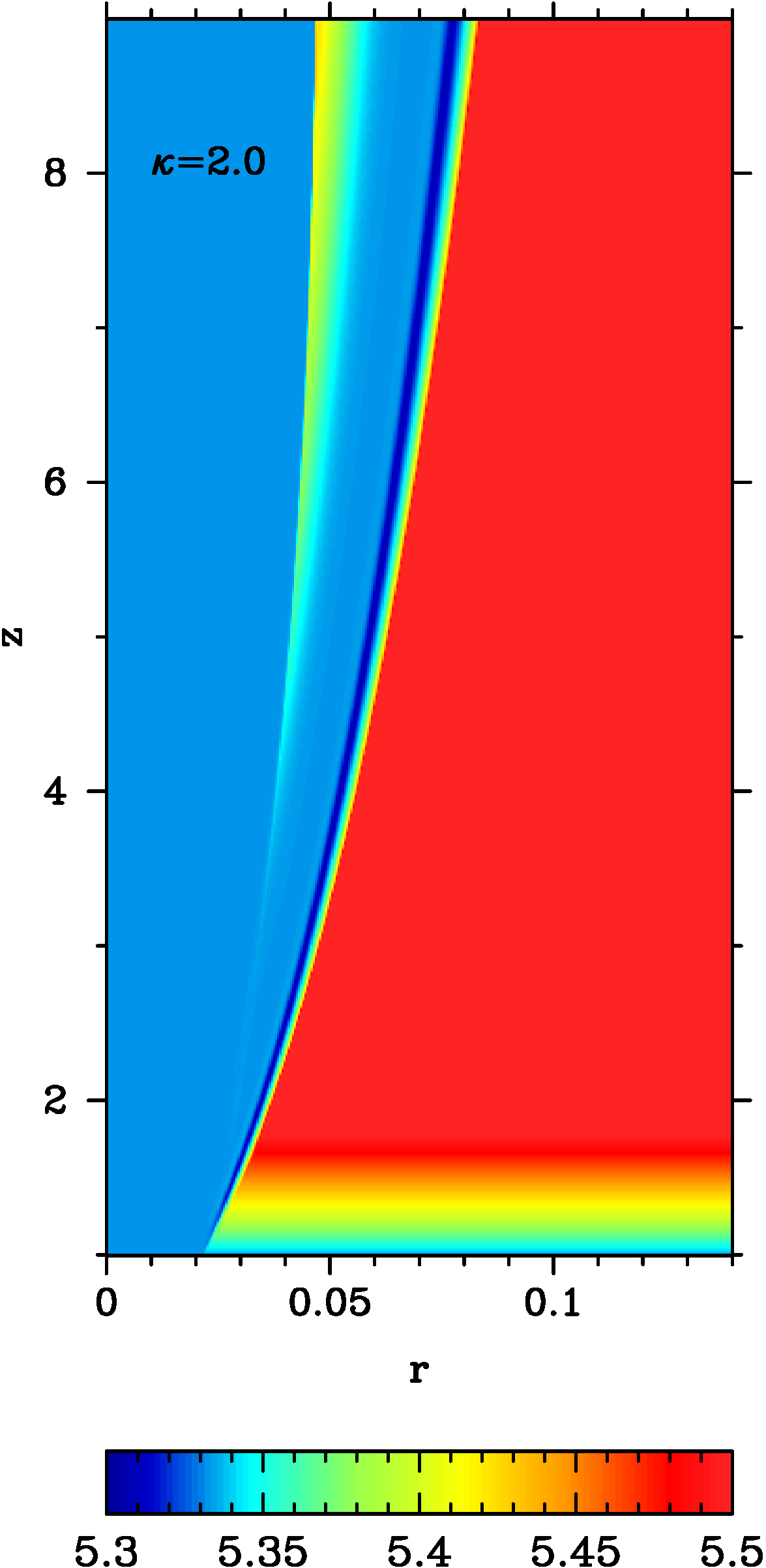}
\caption{
Ultra-relativistically hot jets \citep{KB-12,KB-15} in power-law
atmospheres with $\kappa=8/3, 7/3,$ and 2 (from left to right). The colour-coded images
show $\log_{10}(p\rho^{-\gamma})$. The dark blue region along the jet boundary is 
obviously a numerical artifact as its entropy is lower than that of the initial 
solution anywhere on the grid. 
}
\label{fig:hot2}
\end{center}
\end{figure*}
%

KBB12 studied jets with ultra-relativistic equation of state 
($w=4p$, $\gamma=4/3$ ), propagating in a power-law atmosphere, 

\beq
    p = p_0\fracp{z}{z_0}^{-\kappa}\,.
\label{pla}
\eeq 
These jets emerge from a nozzle at $z=z_0$ with the Lorentz factor $\Gamma_0$, 
opening angle $\theta_0=1/\Gamma_0$ and pressure $p_0$.  The initial velocity distribution 
correspond to a conical flow originating from $z=0$ and hence the initial jet radius 
$r_0=z_0\tan(\theta_0)$. They could only find self-consistent solutions for 
$8/3 \le \kappa <4$ and later argued that for  $\kappa <8/3$ the entropy of the 
shocked layer must increase with the distance along the jet in order for the 
solution to be consistent with the energy conservation \citep{KB-15}.  
They proposed that this additional heating is caused by multiple shocks 
driven into the flow as it gradually collimates. 

We selected the KBB12 model with $\kappa=8/3$ and $\Gamma_0=50$ and made simulations 
on a uniform grid with only 300 cells (each run took only several CPU minutes on a 
laptop using only one core of its processor). Our results
are shown in the first panel of figure~\ref{fig:hot1}, which should be compared
with figure~7 in KBB12.  Again we find a very good agreement between the models --
at $z=9$ we have got the jet radius $r\sub{j}\approx0.114$ and the shock radius 
$r\sub{s}\approx0.07$, whereas in KBB12 $r\sub{j}=0.110$ and $r\sub{s}=0.064$.

In order to understand the difference between the cases with $\kappa>8/3$ and 
$\kappa<8/3$, we also computed models with $\kappa=7/3$ and 2 -- the evolution of the Lorentz 
factor in these models is shown in the second and third panels of figure~\ref{fig:hot1} 
respectively. In these plots we see no evidence of the additional shocks proposed in 
\cite{KB-15}. Neither could we find them in plots of other parameters. However, 
figure~\ref{fig:hot1} suggests that in the models with $\kappa=7/3$ and 2 the reconfinement
shock is much stronger than in the model with $\kappa=8/3$. Moreover, the shock strength is 
increasing with the distance along the jet. As the result, the entropy of the shocked layer 
in the models with $\kappa=7/3$ and 2 is higher and its mean value across the layer is
growing with the distance. This is confirmed in Figure~\ref{fig:hot2}, which shows the 
entropy distribution for these models. 
Since KBB12  assumed isentropy of the flow in the shocked layer, 
this could be the reason why their self-similar model fails for $\kappa<8/3$.  
In contrast, in the model with $\kappa=8/3$ the mean entropy of the layer does remain fairly constant. 
Based on these results, we conclude that the value of $\kappa=8/3$ is not special, but the accuracy 
of the constant-entropy approximation used in KBB12 greatly reduces as $\kappa$ decreases.   

The plots in Figure~\ref{fig:hot2} also reveal a thin layer of 
decreased entropy stretching along the jet boundary. As in this layer the entropy is lower 
than anywhere in the initial solution, this is definitely a numerical artifact. We have checked that 
it becomes less pronounced with increased numerical resolution. Moreover, this layer forms well 
inside the jet and thus its origin is not related to the resetting procedure but is a property of 
our time-dependent code. 

\begin{figure*}
\begin{center}
\includegraphics[height=5cm]{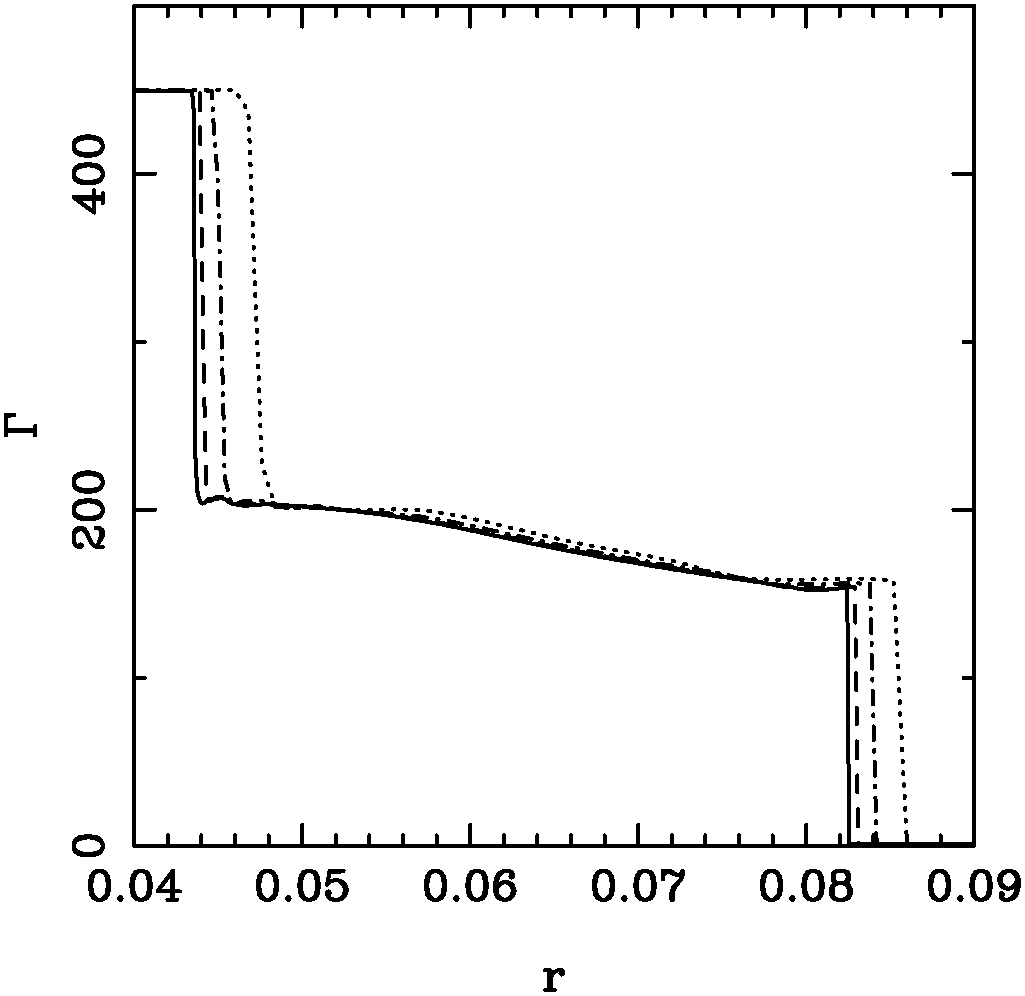}
\includegraphics[height=5.1cm]{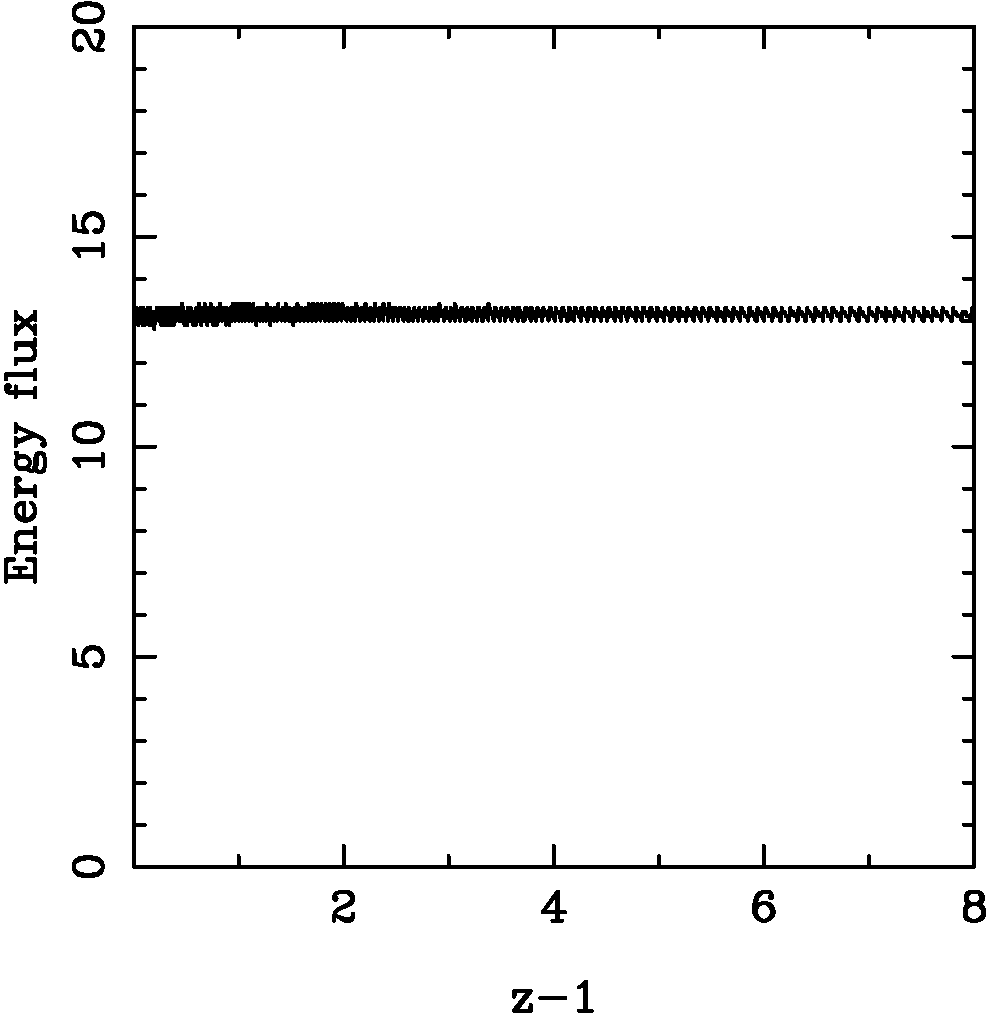}
\caption{
Accuracy of the ultra-relativistic hot jets solution for the model with
atmospheres with $\kappa=2$. The left panel shows the Lorentz factor at $z=9$ 
for models with 150 (dotted), 300 (dot-dashed), 600 (dashed line) and 1200 
(solid line) grid points. The right panel show the total energy flux as a function 
of the distance from the nozzle for the model with 300 grid points.  
}
\label{fig:hot2-conv}
\end{center}
\end{figure*}

{We choose the model with $\kappa=2$, to illustrate the convergence and accuracy of our 
numerical solutions. The left panel of Figure~\ref{fig:hot2-conv} shows the Lorentz factor 
distributions found at $z=9$ for runs with different number of grid cells in the computational 
domain, increasing from 150 to 1200 cells. As one can see, the solutions converge as in a 
first-order accurate scheme. The right panel shows the evolution of the total energy flux
along the jet. It remains fairly constant, as expected for a conserved quantity. As the jet 
boundary jumps from one cell to another, a low level noise is introduced to this integral 
variable. }       

%
%

\subsection{Magnetised jets. 1D versus 2D solutions}
\label{sec:mag-jets}

The steady-state structure of magnetised jets is more complex, mainly due to the 
non-trivial contribution of the magnetic tension to the force balance. 
A number of authors have tackled this problem analytically using various approximations 
\citep[e.g.][]{ZBB-08,lyub-09,lyub-10,KB-12a}. However, none of these studies
deliver a model suitable for detailed testing of our numerical approach. 
Dubal \& Pantano \cite{DP-93} studied the steady-state structure of relativistic jets with
azimuthal magnetic field using the method of characteristics. This would be a good test case but 
the setup of their simulations is ambiguous. We have tried several variants of the setup but 
each time failed to reproduce the results. The mechanisms of magnetic collimation and 
acceleration of relativistic jets were studied numerically by \cite{kbvk-07,
  2009MNRAS.394.1182K} and \cite{tchekhovskoy2010a} using a ``rigid
wall'' outer boundary.   While this allows for a well-controlled experiment, 
Komissarov et al.\cite{kvkb-09}
have shown that the connection between the shape of the boundary and the external pressure 
gradient is not straightforward, with significant degeneracy.       
For this reason, we concluded that in the magnetic case the best way of testing 
the performance of our 1D approach would be via new 2D axisymmetric time-dependent 
simulations using the relativistic AMRVAC code \citep{amrvac,PXHMK-14}. 

\begin{figure*}
\begin{center}
\includegraphics[width=0.7\textwidth]{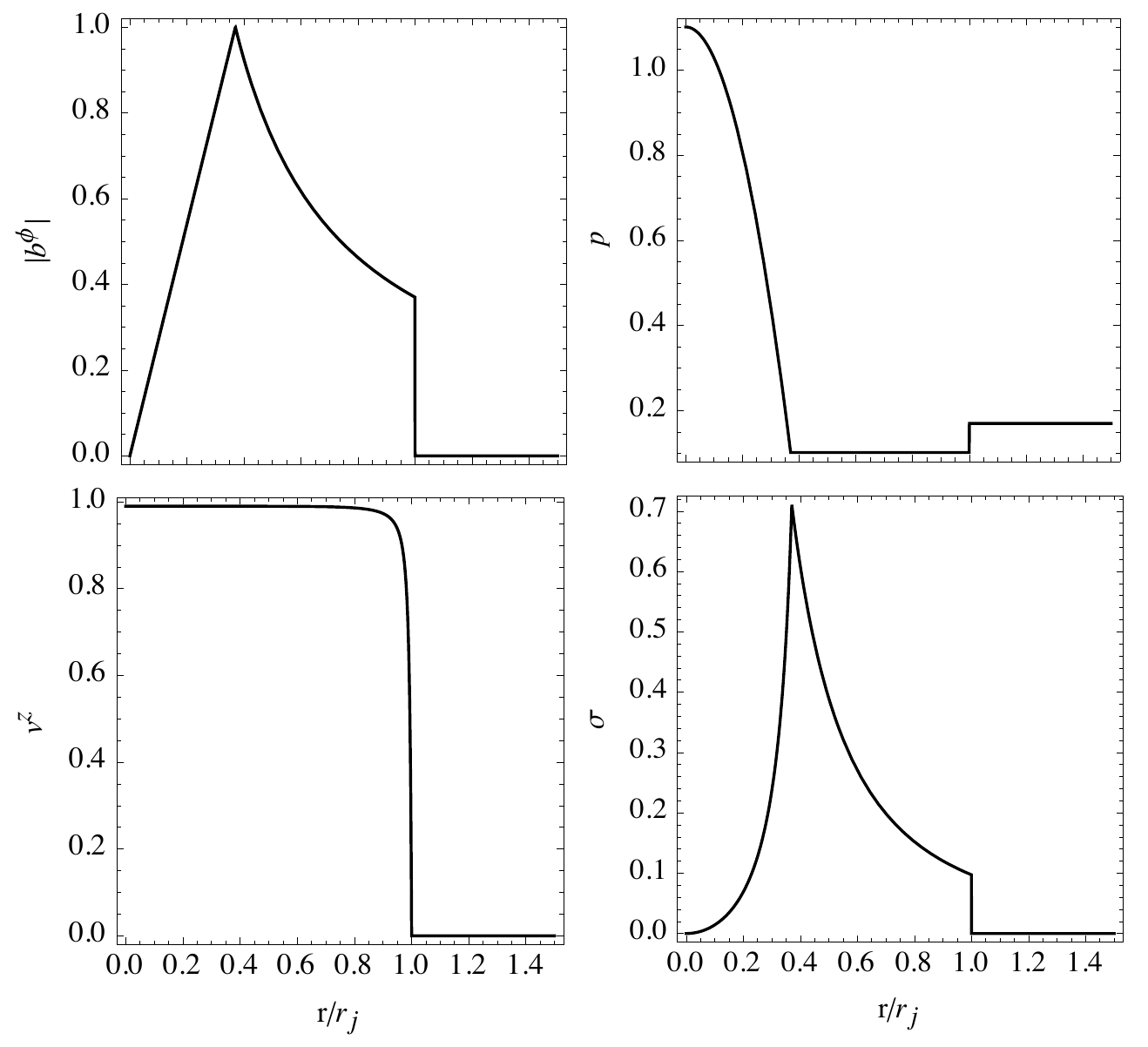}
\caption{Initial radial structure of the magnetised jets in the test simulations described 
in Sec.\ref{sec:mag-jets}.}
\label{fig:cylind}
\end{center}
\end{figure*}

The problem we selected for this test is similar in its setup to the one described in 
Sec.\ref{sec:validation_hot} as it also involves a  jet propagating through 
the atmosphere with the power-law pressure distribution (\ref{pla}), and the nozzle is 
still located at $z=z_0$. However, this time the jet is magnetised and the rest mass density
of its particles is not negligibly small.  
The jet structure at the inlet is that of a cylindrical jet in magnetostatic 
equilibrium, which satisfies the following force balance equation  

\beq
  \oder{p_t}{r} + \frac{b^\phi}{r} \oder{rb^\phi}{r} =0\,,
\label{eq-eqv}     
\eeq
where $b^\phi=B^\phi/\Gamma$ is the azimuthal component of the magnetic field as measured 
in the fluid frame using normalised basis and $p_t$ is the sum of the gas pressure and 
the magnetic pressure due to the axial magnetic field $B_z$ \citep{ssk-mj99}.  
Equation (\ref{eq-eqv}) has infinitely many solutions -- given a particular distribution 
for $b^\phi(r)$ one can solve this equation for the corresponding distribution of the 
pressure $p_t(r)$. We adopted the ``core-envelope'' solution of Komissarov \cite{ssk-mj99}: 
\beq
b^\phi(r) = \left\{
\begin{array}{ccl}
  b_m(r/r_m) &;& r<r_m \\ 
  b_m(r_m/r) &;& r_m<r<r_j
  \\ 0&;& r>r_j
\end{array}
\right. , 
\label{eq:bphi}
\eeq
\beq
p_t(r) = \left\{
\begin{array}{ccl}
  p_0\left[\alpha+\frac{2}{\beta_m}(1-(r/r_m)^{2})\right] &;& r<r_m \\ 
  \alpha p_0 &;& r_m<r<r_j \\ 
  p_0 &;& r>r_j
\end{array}
\right. ,
\label{eq:pressure}
\eeq
where 
\beq
\beta_m= \frac{2 p_0}{b_m^2},\qquad \alpha=1-(1/\beta_m)(r_m/r_j)^2\,, 
\eeq
$r_j$ is the jet radius and $r_m$ is the radius of its core (Note a typo in the expression 
for $\alpha$ in \cite{ssk-mj99}.). As one can see, the core
is pinched and in the envelope the magnetic field is force-free. This may be combined 
with any distribution of density and axial velocity. We imposed $\rho=\rho_0$ and 
\beq
\Gamma(r) = \Gamma_0 \left(1-(r/r_j)^\nu\right) + (r/r_j)^\nu\,, 
\eeq
with $\nu=8$; this gives an almost ``top-hat'' velocity profile. 
The velocity vector is set to be aligned with the jet axis, so $v_r=v_\phi=0$. 
This solution is illustrated in Figure~\ref{fig:cylind}. 

We considered two models, A and B. In the models A, the magnetic field is purely 
azimuthal and the other parameters are $r_j=1$, $r_m=0.37$, $b_m=1$, $\rho_0=1$, 
$z_0=1$, $\beta_m=0.34$, $\Gamma_0=10$. 
The local magnetisation parameter $\sigma=b^2/w$ does not exceed 
$\sigma_{\rm max}=0.7$ in this model and thus the jet is only moderately magnetised. The jet 
core is relativistically hot, with the gas pressure reaching $p_{max}=\rho$ at the 
axis, which opens the possibility of efficient hydrodynamic acceleration once the jet 
is allowed to expand. In the simulations we used the adiabatic equation of state 
$w=\rho+(\gamma/\gamma-1)p$ with $\gamma=4/3$. 

In model B, this configuration is modified to include nonvanishing longitudinal 
magnetic field $B_z$. In particular, we considered the case where  
the gas pressure $p=\alpha p_0$ everywhere within the jet and 
\beq
B^z =
\left\{
\begin{array}{ccl}
 p_0\left[\frac{2}{\beta_m}(1-(r/r_m)^{2})\right] &;& r<r_m\\
 0 &;& r>r_m\,,
\end{array}
\right.
\eeq
which keeps $p_t$ unchanged. 
In this model, the magnetic field is force-free not only in the envelope but also in the core. 
The other parameters of this model that differ from those of model A  are  $\rho_0=0.05$ and 
$\beta_m=0.14$.  The corresponding magnetisation is much higher, with  
$\sigma_{\rm max}=17$.

Model B turned out too stiff for our 2D code, but presented no problems in 1D 
simulations. For this reason we compare here the 1D and 2D results for model A only.
In these simulations we used the atmosphere with $\kappa=1$. 
The computational domain is $20\,r_j$ in the radial direction and $800\,r_j$ 
in the axial direction.       

\begin{figure}
\begin{center}
\includegraphics[width=0.4\textwidth]{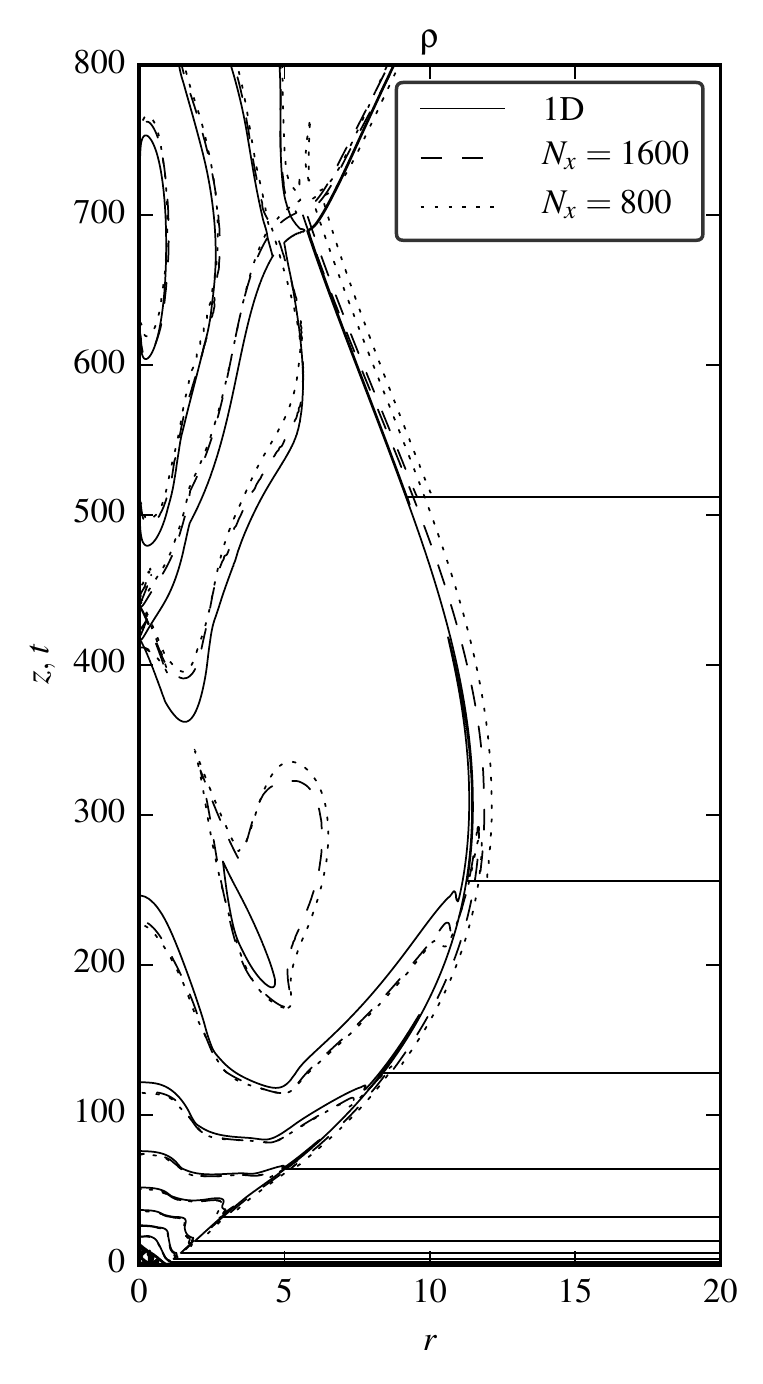}
\caption{Converged 1D solution for a stationary magnetised jet  
  (solid lines) and two corresponding 2D solutions found via the relaxation method, one  
  with the $1600\times3200$-resolution (dashed lines) and one with the
  $800\times1600$-resolution (dotted lines). The lines are 10 rest-frame density 
  contours consecutively spaced by the factor of one half from the 
  starting value of $\rho_{\rm max}=1$.  
  The solution involves a reconfinement shock which reaches the jet 
  axis at $z\approx400$.}
\label{fig:2Dcontours}
\end{center}
\end{figure}

First, let us describe the overall jet structure found in these simulations. 
Initially, as the jet enters the region of rapidly declining external pressure, 
it expands rapidly and a rarefaction wave moves towards its axis.  
Eventually, the jet becomes over-expanded, its expansion slows down, 
and a reconfinement shock sets in. It reaches the axis at $z\approx400$, 
gets reflected and then returns to the jet
boundary at $z\approx700$ (see Figure~\ref{fig:2Dcontours}). 

To quantify the convergence of the 1D simulations we carried out simulations 
with three different resolution and used this data to determine the
grid-convergence index 

\beq \eta \equiv-\ln
\left(\frac{|f_2-f_1|_1}{|f_1-f_0|_1}\right) /\ln{2} 
\eeq 
where $f_1,f_2$ are solutions with doubled and quadrupled resolution
compared to $f_0$ and $|f_a-f_b|_1$ is the difference between two
solutions in the $L_1$ norm. We found that $\eta\approx 1$, as this is 
expected for a TVD scheme in the presence of discontinuities. 
At $6400$ grid cells in the radial direction, the density contours become 
visually unchanged on the scale of figure~\ref{fig:2Dcontours}. The 1D solution 
with this resolution was used for comparison with the results of our 2D simulations. 
In what follows we refer to it as the ``converged'' 1D solution. 

The initial solution in our 2D simulations was constructed via 
interpolation of the converged 1D solution onto the 2D cylindrical grid. 
Since we did not include gravity to balance the pressure gradient in the 
external atmosphere, in order to preserve the atmosphere 
in its initial state the atmospheric parameters were reset to their initial
values every time step, just like this was done in the 1D case.
In order to test the convergence of 2D solutions, we made three 
runs with doubled resolution, $N_r=400$, 800, and 1600 cells in the radial
direction. The number of cell in the axial direction was always twice the number 
of cells in the radial one. 

%
%

Typically, the 2D solutions exhibited some evolution at first but then
quickly settled into a stationary state. For example in the case of
$N_r=400$, the timestep-to-timestep relative variation of the
conserved flow variables dropped below $6\times10^{-6}$ at $t=1000$ and
remained approximately constant thereafter.  Furthermore, the relative $L_1$ error of
density between times $t=1000$ and $t=3000$ was $2.8\times10^{-4}$,
indicating that a stationary state had been reached.  The 2D
solutions converge with the grid-convergence index $\eta>1.25$ over the
entire simulated time.

The difference between the converged 1D solution and the relaxed 2D solutions with 
with $N_r=800$ (dotted lines) and $N_r=1600$ (dashed lines) is illustrated 
in Figure~\ref{fig:2Dcontours} which shows the mass density distribution.    
One can see that the 2D solutions are very close to
the 1D solution and that the difference decreases with the resolution
of 2D runs.  To quantify the difference between the relaxed 2D
solutions and the converged 1D approximate solution we introduce the
parameter
\beq 
\delta\rho=|\rho_{2D}-\rho_{1D}|_1/\langle\rho_{1D}\rangle \, .
\eeq 
For the 2D solution with $N_r=400$ cells in the radial direction  
we obtain $\delta\rho\simeq6\%$, for $N_r=800$  $\delta\rho\simeq4.3\%$
and for $N_r=1600$ the relative error decreased to $\delta\rho\simeq3.2\%$.  
This shows that the approximation error of our 1D approach is at the level of no more 
than $3\%$.

\subsection{Magnetised jets in power-law atmospheres}
\label{sec:powerlaw1}

Komissarov et al.\cite{kvkb-09} derived an approximate equation for the radius of 
highly magnetised jets, in the limit where it strongly exceeds that of the light 
cylinder. Using this equation they concluded that in the case of power-law 
atmosphere with $0<\kappa<2$ the jet radius increases as 

\beq 
   r_j \propto z^{\kappa/4}\,. 
\label{eq:rj-kappa}
\eeq 
Lyubarsky \cite{lyub-09} developed the theory of Poynting-dominated
jets further and using more accurate analysis found that the expansion
is modulated by oscillations with the wavelength growing as

\beq 
   \lambda \propto z^{\kappa/2}. 
\label{eq:lambda-kappa}
\eeq  
These oscillations can be understood as a standing magneto-sonic wave bouncing across 
the jet. Indeed, denote the wave speed as $a_m$. Then the jet crossing time is 
$\tau_c=r_j/a_m$ in the co-moving jet frame and $t_c=\Gamma\tau_c$ in the rest frame 
of the atmosphere. As the wave is advected along the jet almost at the speed of light 
the wavelength of the associated structure is  

\beq
\lambda \simeq \Gamma \frac{c}{a_m} r_j \,.
\label{eq:lambda-basic}
\eeq 
Since for the jets considered in \cite{kvkb-09,lyub-09} $a_m\simeq c$ and 
$\Gamma \propto r_j$ we obtain $\lambda\propto r_j^2$ and using Eq.\ref{eq:rj-kappa} 
recover Eq.\ref{eq:lambda-kappa}.   
The results (\ref{eq:rj-kappa},\ref{eq:lambda-kappa}) are well suited for testing 
of our approach. To this aim, we carried out additional 1D simulations with models 
A and B  described in Sec.\ref{sec:mag-jets}. 

%
\begin{figure}
\includegraphics[height=9cm]{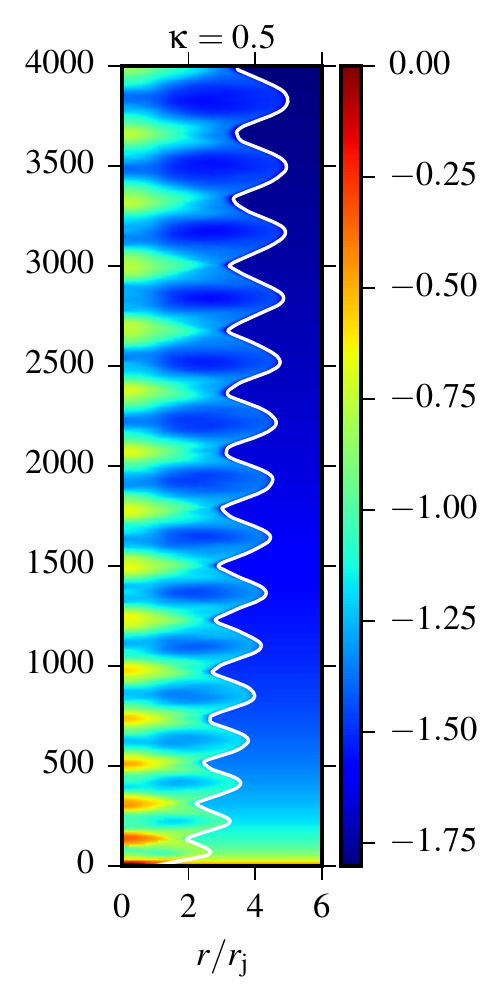}
\includegraphics[height=9cm]{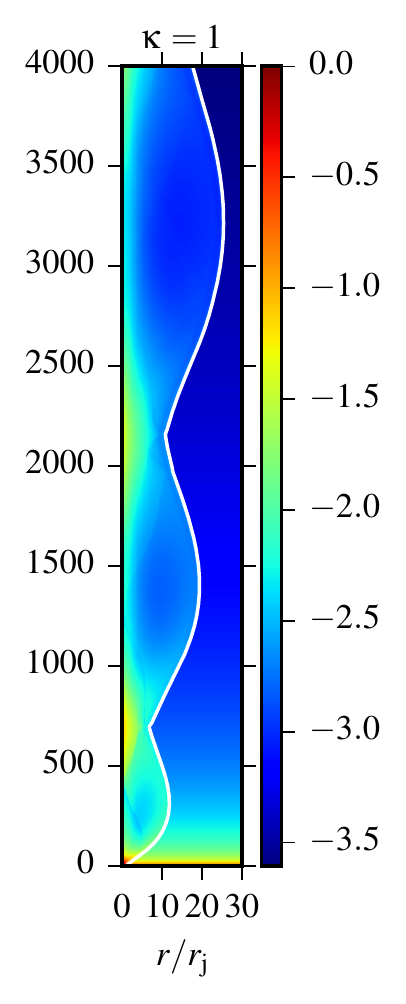}
\caption{Structure of steady-state magnetised jets obtained via time-dependent 1D
  simulations.  The plots show the co-moving density distribution for
  model A with $\kappa=0.5$ and $\kappa=1$. The distance along the
  vertical axis is defined as $z=ct/r_j$, where $r_j$ is the initial
  jet radius. The white contour shows the jet boundary, located using
  the passive scalar. }
\label{fig:1d}
\end{figure}
%

Since in model A the jet is not 
Poynting-dominated, it allows us to explore the regime not covered in \cite{lyub-09}.  
To see how sensitive these results may be to the assumptions made in 
\cite{kvkb-09,lyub-09} let us consider unmagnetised relativistic jets. From the mass 
conservation law we obtain $r_j \propto (\Gamma \rho)^{-2}$. For relativistically 
cold jets with $p\ll \rho c^2$ we have $\Gamma\simeq\mbox{const}$ and thus 

\beq
r_j \propto z^{\kappa/2\gamma} \,,
\label{eq:r_cold}
\eeq
whereas for the hot jets $\Gamma\propto r_j$ and thus 
\beq
r_j \propto z^{\kappa/4} \,,
\label{eq:r_hot}
\eeq
where we put $\gamma=4/3$. The last result is the same as for the Poynting-dominated 
jets. Even for the cold jets the difference is rather minor, e.g. for $\gamma=5/3$ 
the index in Eq.\ref{eq:r_cold} differs from $\kappa/4$ only by $\kappa/20$ and 
for $\gamma=4/3$ by $\kappa/8$.

In order find $\lambda$ we note that for cold jets 
$a_m^2 \propto (p/\rho) \propto z^{-\kappa(1-1/\gamma)}$ and hence Eq.\ref{eq:lambda-basic}
yields Eq.\ref{eq:lambda-kappa} independently of the value of $\gamma$. For hot jets, 
$a_m \simeq\mbox{const}$ and Eq.\ref{eq:lambda-basic} still 
leads to Eq.\ref{eq:lambda-kappa} if we use $\gamma=4/3$. 
Thus, the law (\ref{eq:lambda-kappa})  for the wavelength of oscillations is very robust.

\begin{figure*}
\includegraphics[height=10cm]{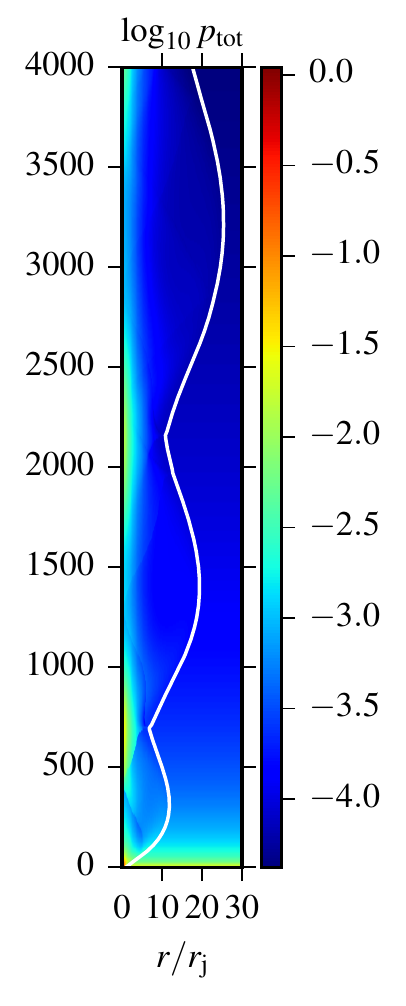}
\includegraphics[height=10cm]{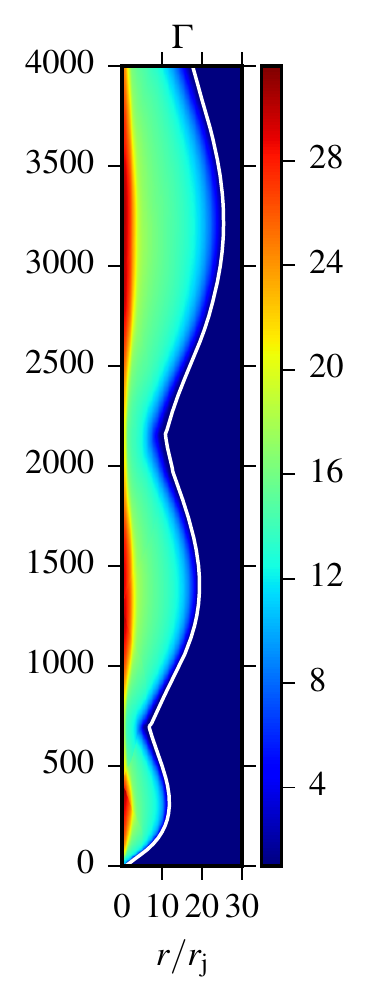}
\includegraphics[height=10cm]{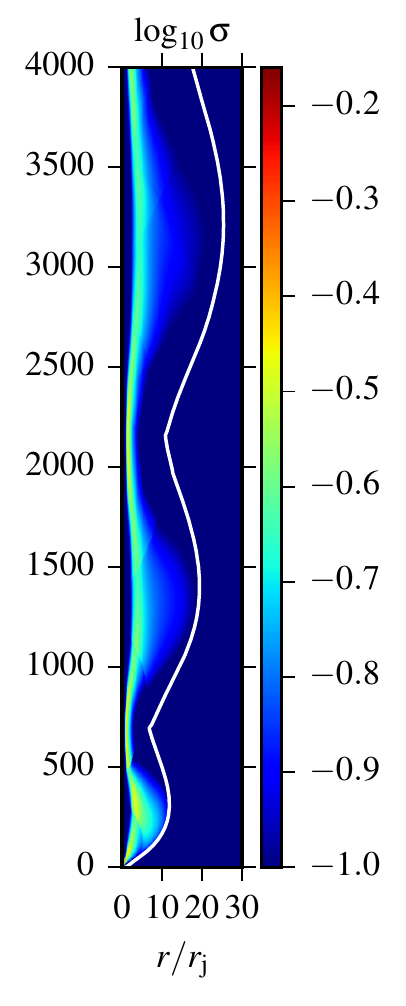}
\caption{Structure of the steady-state magnetised jet in model A 
with $\kappa=1$, obtained via time-dependent 1D simulations. 
From left to right, the plots show the total pressure,
Lorentz factor and the magnetisation parameter $\sigma$. 
Jet oscillations cause compression in the
squeezed regions as well as re-acceleration of the bulk flow as the
flow expands.  The majority of acceleration occurs in the thermally
dominated core.  A reconfinement shock is clearly visible in the total
pressure and magnetisation plots.  }
\label{fig:model_a}
\end{figure*}

Figure~\ref{fig:1d} illustrates the overall jet structure in model A and its 
response to changes in the parameter $\kappa$ of the external atmosphere. 
One can see that this weakly magnetised jet also shows a combination of secular 
expansion and oscillations. These oscillations appear to be a generic feature of the 
adjustment process of supersonic jets to variations of external pressure, which occurs by 
means of magneto-sonic waves travelling across the jet.  
In the very beginning, the decrease of external pressure makes the jet 
under-expanded and a rarefaction wave is launched from the jet boundary towards 
the jet axis. Behind this wave the radial velocity is positive and the flow 
expands. The rarefaction reduces the jet pressure and at some point it becomes 
over-expanded. Now a compression wave is driven inside the jet. Behind this 
wave the jet expansion slows down and eventually turns into a contraction. 
The contraction increases the jet pressure and at some point it becomes 
under-expanded again and then the whole cycle repeats. 

\begin{figure*}
\includegraphics[height=10cm]{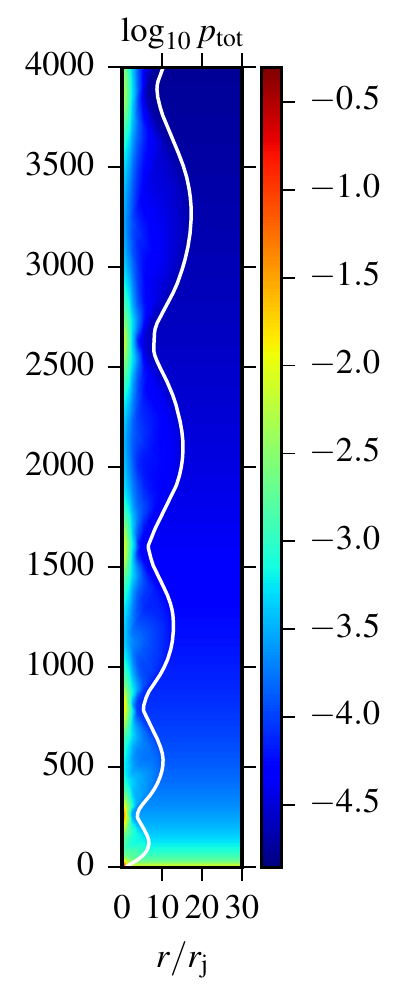}
\includegraphics[height=10cm]{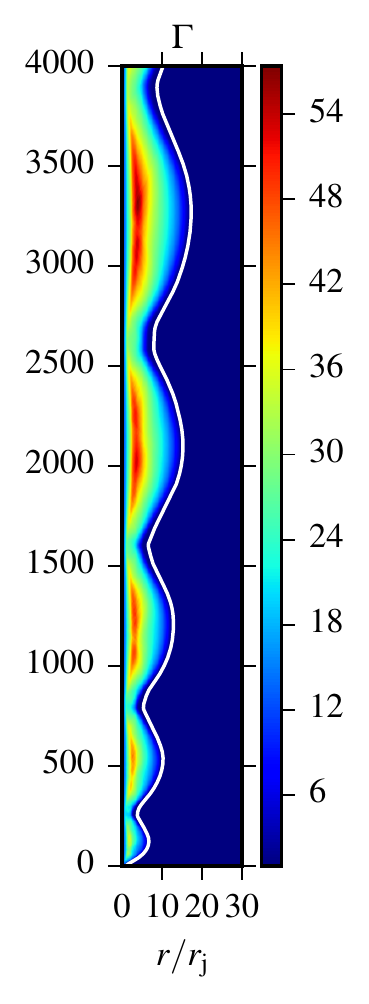}
\includegraphics[height=10cm]{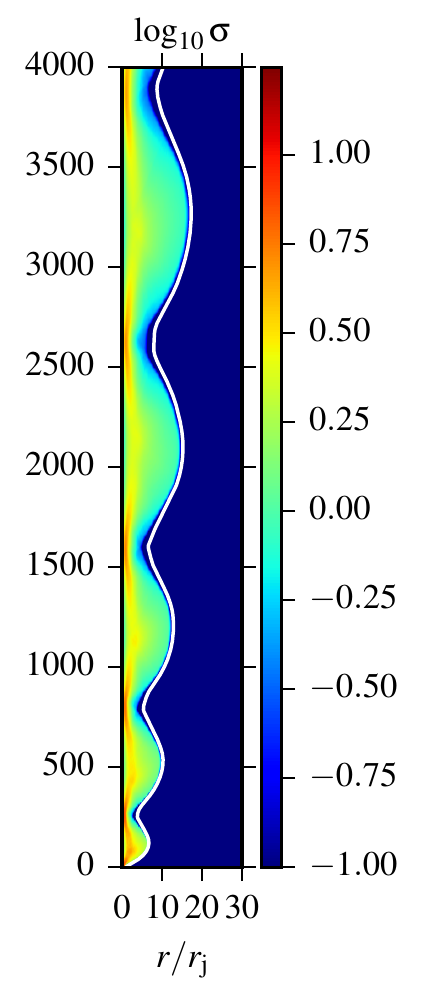}
\caption{As figure \ref{fig:model_a} for model B with $\kappa=1$.  As the
  Poynting-flux vanishes on the axis (and the thermal energy is
  negligible), we obtain a hollow jet with fastest region away from
  the axis.  Due to the increased fast-magneto-sonic speed (thus lower
  Mach-number) compared to the case of figure \ref{fig:model_a}, no
  reconfinement-shock forms and the jet-oscillation frequency is
  increased.}
\label{fig:model_b}
\end{figure*}

The deviation from 
the force-balance corresponding to the secular jet expansion is due to the 
finite propagation speed of the waves - as they move across the jet 
they are also advected downstream by the supersonic flow. As the result, the jet 
interior reacts to the changes in the external pressure with a delay. It keeps 
expanding when the internal pressure is already too low and keeps contracting 
when it is already too high.     
As $\kappa$ increases, the wavelength of the oscillation increases as well. 
This is expected as the more rapid overall expansion of the jet in an atmosphere with 
larger $\kappa$ means that it takes longer for a magneto-sonic wave to traverse the jet, 
not only as the result of the larger jet radius but also as the result of its higher 
Mach number (and hence smaller Mach angle).  

Overall, this is very similar to the well-known evolution of under-expanded 
supersonic jets studied in laboratories. Normally, their compressive transverse waves  
steepen into shocks. In our model A with $\kappa=1$ we also detect shocks, 
but they become progressively weaker, suggesting that they may disappear further 
out along the jet. For $\kappa=0.5$, shocks do not form at all. The exact reason 
for this in not yet clear.   

\begin{figure*}
\includegraphics[width=6cm]{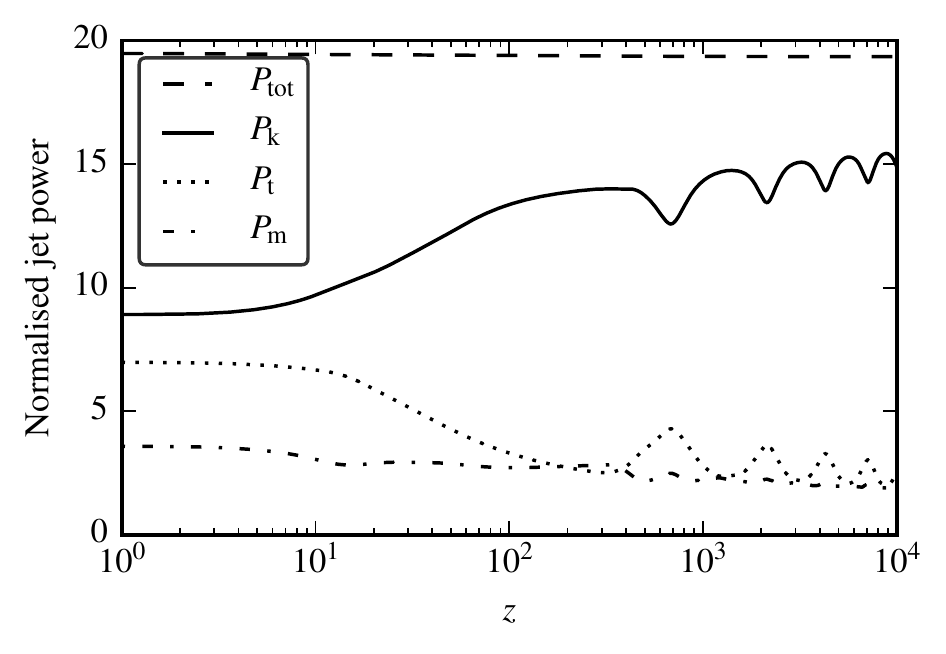}
\includegraphics[width=6cm]{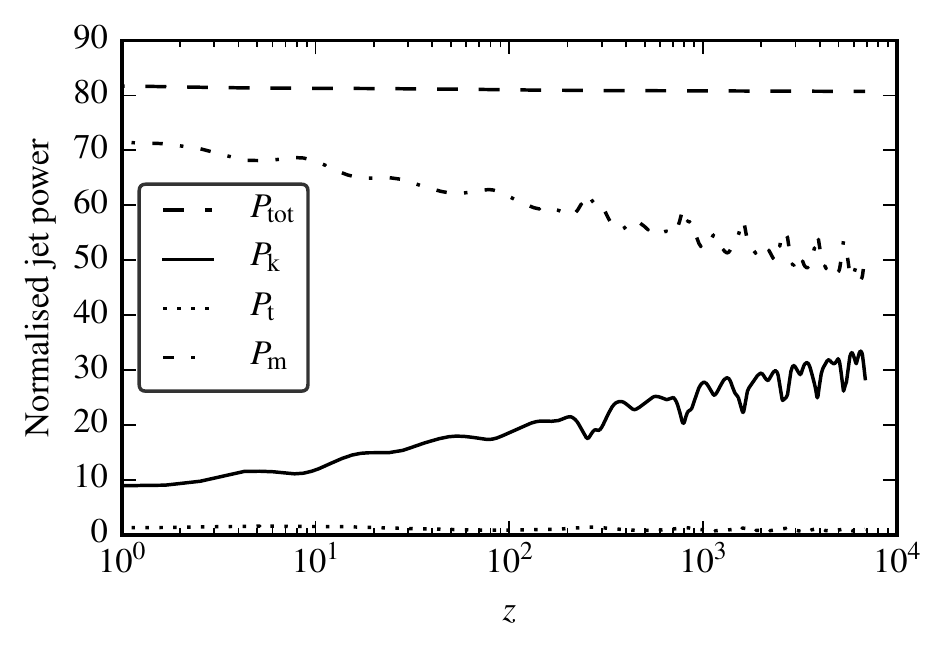}
\caption{Evolution of energy fluxes with the distance along the jet 
in models A (left panel) and B (right panel).
The curves show fluxes of the total (dash), kinetic (solid), thermal (dot) and 
magnetic (dot-dash) energies. Each is normalised to the rest-mass flux 
}
\label{fig:energies}
\end{figure*}

Figure~\ref{fig:model_a} shows the evolution of other flow  parameters in model A 
with $\kappa=1$.  Both the secular and oscillatory behaviour of the jet 
radius are mirrored in the variation of the Lorentz factor. 
The secular expansion leads to secular increase of the Lorentz factor as both 
the thermal and the magnetic energy are converted into the kinetic energy 
of the flow. The thermal acceleration is most pronounced in the jet core, which is 
relativistically hot at the inlet.  
The oscillations of the jet radius lead to additional increase of 
the Lorentz factor during the expansion phase and its decrease upon contraction. 
{The left panel of Figure~\ref{fig:energies} shows the dynamics of energy 
fluxes for this jet. These are found via integration over the jet cross-section 
of $b^2 \Gamma^2 v^z -b^0b^z $ for the magnetic energy, $ \rho\Gamma^2 v^z$
for the kinetic energy and $(w-\rho)\Gamma^2 v^z$ for the thermal energy. 
The results are normalised to the rest-mass flux, obtained via integration 
of $\rho\Gamma v^z$. As the result of this normalisation, the kinetic 
energy flux has the meaning of mean actual Lorentz factor of the jet, whereas for 
the thermal energy and magnetic energies these are the gains in the Lorentz factor, 
which can be achieved upon full conversion of these energies into the kinetic one. 
The main feature of the plot is a conversion of the thermal 
energy into the kinetic one (the magnetic energy is highly sub-dominant from the 
start). This conversion is largely completed during the initial phase of monotonic
expansion, which  lasts up to $z=200$.   In the second phase, the thermal 
energy flux is comparable to the magnetic, and they are being converted to the 
kinetic energy at more of less the same and rather slow rate.       
Strong oscillations are superimposed upon this secular evolution, with the kinetic
(thermal) energy reaching local maxima (minima) at the locations of jet bulging.}   
   
Figure~\ref{fig:model_b} shows the same parameters for the highly magnetised jet of
model B with $\kappa=1$. In this model, the reconfinement shock is no longer present. 
This may be related to the fact that in this model the fast magneto-sonic speed is 
higher and the corresponding jet Mach number is lower, at the inlet $M\simeq3$ compared 
to $M\simeq10$ in model A. The lower Mach number is also responsible for 
the observed lower wavelength of the jet oscillations as it takes less time for the 
waves to traverse the jet. {In this model, the jet is magnetically-dominated and 
the main feature of its energy balance is a gradual conversion of the magnetic energy 
into the kinetic one (see the right panel of Figure~\ref{fig:energies}).}

\begin{figure}
\includegraphics[width=0.80\textwidth]{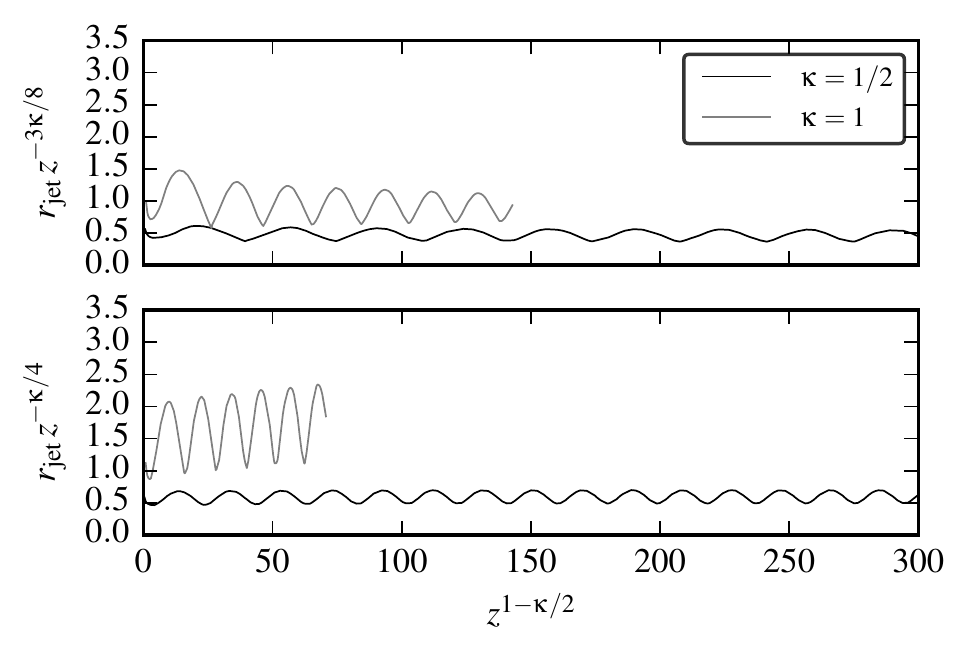}
\caption{Compensated jet-expansion laws for models A 
  (\textit{top}) and B (\textit{bottom}).
  In both models the expected average expansion is captured quite well.  
  To show that the oscillation wavelength scales as $\lambda\propto z^{\kappa/2}$,
  the x-axis has been rescaled accordingly.  In order to visually
  separate the curves corresponding to different values of $\kappa$, 
  they have been shifted up by a factor of $\kappa$.}
\label{fig:expansion}
\end{figure}

The theoretical predictions for the secular evolution of the jet radius 
and the wavelength of its oscillations are put to a quantitative test 
in figure \ref{fig:expansion}, which shows the jet radius rescaled according 
to its expected secular evolution against $z^{1-\kappa/2}$. In such plots, 
the mean jet radius and the wavelength of oscillations should remain constant. 
For the highly magnetised jet of model B the scaling factor is $z^{\kappa/4}$ 
and for the low magnetised jet of model A it is $z^{3\kappa/8}$, as
appropriate for a cold hydrodynamic jet with $\gamma=4/3$.  In
general, we obtain a very good agreement with the theoretical scalings
for the mean jet radius, both in the low- and high-magnetisation
limit.  A small departure from the $z^{\kappa/4}$-scaling is observed for
case B with $\kappa=1$ -- it expands slightly faster.  This could be 
because the jet magnetisation is not sufficiently high and decreases 
more rapidly with distance than in the atmosphere  with $\kappa=0.5$. 
The evolution of the wavelength scaling is also in a very good agreement with 
the theory - the residual error is between $0.7\%$ and $3.4\%$.
%

\section{Conclusions}
\label{sec:concl}

In this paper we presented a novel numerical approach, which can be used 
to determine the structure of steady-state relativistic jets. It is based 
on the similarity between the two-dimensional steady-state equations and 
the one-dimensional time-dependent equations of SRMHD with the cylindrical 
symmetry in problems involving narrow highly-relativistic ($v_z\approx c$) flows. 
Such similarity has already been utilised in the so-called ``frozen pulse'' 
approximation where dynamics of time-dependent relativistic flows is analysed 
using the steady-state equations \cite{PN-93,VK-03a,SV-13}.
Here we do the opposite and construct approximate steady-state solutions via 
numerical integration of the time-dependent equations. The main advantage of this 
approach is utilitarian. First, it allows us to use computer codes for relativistic 
MHD (or hydrodynamics in the case of unmagnetised flows), which are now widely 
available, in place of highly-specialised codes for integrating steady-state 
equations, which are not openly available at the moment. Moreover, the reduced 
dimensionality means that the computational facilities can be very modest -- 
a basic laptop will suffice. In contrast, the relaxation method based on 
integration of two-dimensional time-dependent equations can be computationally quite 
expensive.  

We compared numerical solutions obtained with this approach with analytical models 
and numerical solutions obtained with other techniques.  The considered problems 
involved a variety of flows both magnetised and unmagnetised, with different equations 
of state and external conditions.  The results show that the method is sufficiently 
accurate and robust.                          

Although we focused only on relativistic flows, we see no reason why this 
approach cannot be applied to non-relativistic hypersonic flows. For such flows, 
the axial velocity of bulk motion plays the role of the speed of light in the 
substitution $z=ct$ used in our derivations. 

As a byproduct of our test simulations, we obtained two results of astrophysical 
interest. We demonstrated that the failure of the self-similar model of the 
jet reconfinement in power-law atmospheres with the index $\kappa<8/3$ 
\cite{KB-12} is rooted in the assumption of isentropy of the shocked layer, 
which is made in this model. In reality, the reconfinement shock becomes stronger 
with the distance along the jet, resulting in a strong spatial variation of 
the entropy. We also found that the radial oscillations of steady-state jets, 
discovered in the analytical models of Poynting-dominated jets \cite{lyub-09} 
is a generic part of the jet adjustment to the space-variable external pressure 
and not specific to the high-magnetisation regime only. The oscillations are 
standing waves induced by the variation.           

The steady-state solutions are useful for elucidating some key factors in flow 
dynamics and may closely describe some of the observed phenomena in astrophysical 
jets. However, they are often subject to various instabilities which may dramatically
modify the flow properties. Most instability studies, both analytical and numerical, deal 
with very simple problems where the steady-state solution is readily available. 
In more realistic setup, the issue of finding the steady-state solution, which can then 
be subjected to perturbations, becomes more involved and this is where our method can 
be applied in the instability studies.


\begin{backmatter}

\section*{Competing interests}
The authors declare that they have no competing interests.

\section*{Author's contributions}
SSK and ML carried out the work on the justification of the numerical approach.
OP and SSK carried out the test simulations. All authors contributed to
writing the manuscript.

\section{Acknowledgments}
SSK and OP were supported by STFC under the standard grant ST/I001816/1. 
SSK and ML were supported by NASA under the grant NNX12AF92G. 
OP thanks Purdue University for hospitality during his visits in 2014.  
We thank the anonymous reviewers for constructive comments and suggestions. 

\bibliographystyle{bmc-mathphys}

\bibliography{BibFiles/mypapers,BibFiles/lyubarsky,BibFiles/lyutikov,BibFiles/jets,BibFiles/hea,BibFiles/numerics,BibFiles/mix,BibFiles/astro}


\end{backmatter}

\end{document}